\documentclass[12pt]{article}
\pdfoutput=1

\usepackage{jstyle}
\usepackage{tikz}
\usepackage{amsthm,mathrsfs,stmaryrd,bm}
\usepackage[vcentermath]{youngtab}
\usepackage{simplewick}
\usepackage{framed,pifont}

\usepackage{diagbox}
\usepackage{tikz}

\definecolor{myred}{rgb}{1.0, 0.45, 0.400}

\usepackage{graphics}
\usepackage{graphicx}
\usepackage{epic}
\usepackage{eepic}

\usepackage{ulem}

\def\a{\alpha}
\def\b{\beta}
\def\g{\gamma}

\def\d{\delta}
\def\D{\Delta}
\def\e{\epsilon}

\def\h{\eta}

\def\l{\lambda}
\def\L{\Lambda}
\def\m{\mu}
\def\n{\nu}

\def\r{\rho}
\def\s{\sigma}

\def\f{\phi}
\def\F{\Phi}
\def\vf{\varphi}

\def\o{\omega}

\usepackage{latexsym}
\usepackage{amsmath}
\usepackage{wasysym}
\usepackage{amsfonts}
\usepackage{pifont}
\usepackage{bm}
\usepackage{epsf}
\usepackage{epsfig,graphicx}
\usepackage{amsmath,amssymb,bm,epsf,epsfig,graphicx}
\tikzstyle{decision}=[diamond, draw,fill=blue!50]
\tikzstyle{line}=[draw,-latex']

\usepackage{diagbox}

\usepackage{color}
\definecolor{rougef}{rgb}{0.56,0,0}
\definecolor{vertf}{rgb}{0,0.5,0}
\definecolor{bleuf}{rgb}{0,0,0.8}
\definecolor{violetf}{rgb}{0.5,0,0.5}

\usepackage[vcentermath]{youngtab}

 \csname
@addtoreset\endcsname{equation}{section}

\def\pe{\prime}
\def\3s{{s \choose 3}}
\def\4s{{s \choose 4}}
\def\5s{{s \choose 5}}
\def\6s{{s \choose 6}}

\def\12{\dfrac{1}{2}}
\def\fr{\frac}
\def\ft{\footnote}

\def\nn{\nonumber}

\def\2{\ell_2}

\def\pr{\partial}

\def\prd{\partial \cdot}

\def\q{\quad}

\def\be{\begin{equation}}
\def\ee{\end{equation}}
\def\bea{\begin{eqnarray}}
\def\eea{\end{eqnarray}}
\def\ba{\begin{array}}
\def\ea{\end{array}}
\def\bec{\begin{center}}
\def\ec{\end{center}}

\def\a{\alpha} 
\def\b{\beta}  
\def\g{\gamma} 

\def\d{\delta} 
\def\D{\Delta}
\def\e{\epsilon}

\def\h{\eta}

\def\l{\lambda}
\def\L{\Lambda}
\def\m{\mu}
\def\n{\nu}

\def\r{\rho}
\def\s{\sigma}

\def\f{\phi}
\def\F{\Phi}
\def\vf{\varphi}

\def\o{\omega}

\def\cA{{\cal A}}
\def\cB{{\cal B}}
\def\cC{{\cal C}}
\def\cD{{\cal D}}

\def\cF{{\cal F}}

\def\cL{{\cal L}}
\def\cM{{\cal M}}

\def\cO{{\cal O}}

\def\cS{{\cal S}}
\def\cT{{\cal T}}
\def\cV{{\cal V}}

\def\lag{\mathcal{L}}

\def\mezzi{\dfrac{1}{2}}
\def\cQ{{\cal Q}}

\author{Dario Francia$^{\, a}$,}
\affiliation{$^a$Scuola Normale Superiore and INFN, Piazza dei Cavalieri, 7 I-56126 Pisa, Italy}
\author{Gabriele Lo Monaco$^{\, b, c}$ and}
\affiliation{$^b$Dipartimento di Fisica, Universit\`a di Pisa, Piazza Fibonacci, 3, I-56126, Pisa, Italy, \\
$^c$Dipartimento di Fisica, Universit\`a di Milano--Bicocca, Piazza della Scienza 3, I-20126 Milano, Italy}
\author{Karapet Mkrtchyan$^{\, d}$}
\affiliation{$^d$Max Planck Institut f\"ur Gravitationsphysik, Am M\"uhlenberg 1, Potsdam 14476, Germany}

\emailAdd{dario.francia@sns.it \\ \hskip 45pt g.lomonaco1@campus.unimib.it \\ \hskip 45pt karapet.mkrtchyan@aei.mpg.de}

\title{\centering
\LARGE{Cubic interactions of Maxwell-like higher spins}}

\abstract{We study the cubic vertices for Maxwell-like higher-spins in flat and (A)dS background spaces of any dimension. Reducibility of their free spectra implies that a single cubic vertex involving any three fields subsumes a number of couplings among different particles of various spins. The resulting vertices do not involve traces of the fields and in this sense are simpler than their Fronsdal counterparts.  We propose an extension of both the free theory and of its cubic deformation to a more general class of partially reducible systems, that one can obtain from the original theory upon imposing trace constraints of various orders. The key to our results is a version of the Noether procedure allowing to systematically account for the deformations of  the transversality conditions to be imposed on the gauge parameters at the free level.}

\begin{document}

\maketitle

\tableofcontents
\newpage

\section{Introduction} \label{sec: intro}

 In this work we construct the cubic vertices on flat and (A)dS backgrounds deforming the free Lagrangians for massless higher-spin fields proposed in \cite{ML}. Their equations of motion are based on a kinetic tensor retaining the same form as the Maxwell tensor for spin-one fields, 
\be \label{Ml}
M \, = \, \Box \, \vf \, - \, \pr \, \prd \vf \, = \, 0 \, ,
\ee
where $\vf$ denotes a rank$-s$ tensor, and rely on an Abelian gauge symmetry with transverse gauge parameters
\be \label{transversality}
\d \, \vf \,  = \, \pr \, \e, \hskip 2cm \prd \e \, =\,  0 \, .
\ee
They propagate a reducible spectrum of massless particles with spin $s, s-2, s-4, \ldots \, $, and it is mainly in this sense that this Maxwell-like description differs from the Fronsdal one \cite{fronsdal}, whose equations describe the degrees of freedom of a single massless particle of spin $s$.  As a consequence,
a given cubic vertex involving a specific triple of Maxwell-like tensors with ranks $s_1, s_2$ and $s_3$, actually subsumes a number of cross-interactions among all the particles with different spins actually carried by each tensor. 

Cubic interactions involving massless particles of arbitrary spins have been investigated from several perspectives, starting from the light-cone results of the G\"oteborg group \cite{Bengtsson:1983pd,Bengtsson:1986kh}, that also provided the very first non-trivial instances of higher-spin interactions ever proposed. The systematics of covariant constructions, together with some explicit instances of couplings, were first discussed in  flat space in \cite{Berends:1984wp, Berends:1984rq}, while with the seminal work of Fradkin and Vasiliev \cite{Fradkin:1986qy} the relevance of (A)dS background was appreciated for the first time.  Subsequent extensive explorations have been performed, culminating in a classification of cubic vertices for massless symmetric higher-spin fields in arbitrary dimensions, both in the light-cone gauge \cite{Metsaev:2005ar,Metsaev:2007rn} and in covariant form \cite{Manvelyan:2010jr} for the case of flat backgrounds. For alternative perspectives and  additional references, as well as for generalisations to (A)dS spaces, see \cite{Bengtsson:1987jt,Fradkin:1991iy,Bekaert:2005jf,Bekaert:2006us,Buchbinder:2006eq,Boulanger:2006gr,
Fotopoulos:2007nm,Fotopoulos:2007yq,Fotopoulos:2008ka,Boulanger:2008tg,Bekaert:2009ud,
Zinoviev:2008ck,Manvelyan:2009tf,Manvelyan:2009vy,Polyakov:2009pk, Manvelyan:2010wp,Taronna:2010qq,
Polyakov:2010qs,Sagnotti:2010at,Zinoviev:2010cr,Fotopoulos:2010ay,Manvelyan:2010je,Mkrtchyan:2010pp,Polyakov:2010sk,
Mkrtchyan:2011uh,Taronna:2011kt,Vasilev:2011xf,Joung:2011ww,Joung:2012rv,Metsaev:2012uy,
Henneaux:2012wg,Joung:2012fv,Manvelyan:2012ww,Buchbinder:2012xa,Joung:2012hz,Boulanger:2012dx,
Cortese:2013lda,Henneaux:2013gba,Joung:2013nma, Bekaert:2015tva,Bengtsson:2016alt,Bengtsson:2016hss}. 

There are three main aspects of our investigation where the differences between Maxwell-like fields and Fronsdal fields are more remarkable:
\begin{itemize}
 \item the structure of the cubic vertex for Maxwell-like fields is simpler than in the Fronsdal case;
 \item the transversality conditions on the free gauge parameters are to be corrected by field-dependent terms. This leads to modifications of the Noether procedure where new equations have to be taken into account;
 \item in the Maxwell-like setting, we argue that the same Lagrangian can describe spectra of various degrees of complexity, upon imposing trace constraints of increasing strength on both fields and parameters. Our construction holds for all these cases, thus providing cubic interactions for putative complete theories with different particle contents.
 \end{itemize}
 
Since the detailed analysis of these issues is somewhat technical, in this introduction we aim to illustrate them in qualitative terms, thus providing a general summary of our results. 
 
\subsection*{Structure of Maxwell-like cubic vertices} \label{sec: cubic_gen}

In the context of interactions of massless fields, flat space is singled out among constant-curvature backgrounds. In flat space, vertices containing a different number of derivatives are essentially independent, as far as gauge invariance is concerned. Therefore, on Minkowski backgrounds, the number of derivatives proves to be a useful guiding marker for classifying cubic interactions  for irreducible massless (Fronsdal) fields. Reducible models are not different in this respect, so that we can study their cubic interactions assuming fixed the overall number of derivatives.

Thus, as far as its essential structure is concerned, any cubic vertex can be written as follows:
\be
\cL_1 \, \sim \, \pr^{\, n} \, \vf_1 \, \vf_2 \, \vf_3\, , 
\ee
involving a total of $n$ derivatives, acting in a prescribed way on the tensors $\vf_i$ of rank $s_i,\,  i=1, 2, 3$. As customary in these types of problems, one starts from an ansatz for $\cL_1$ where no divergences or traces of the fields $\vf_i$ are taken into account (the so-called transverse-traceless, or  briefly TT, sector), to then proceed to include them in order to set up a proper scheme of cancellation of the gauge variation at each step. The possible types of terms are summarised in Table \ref{table:1}, with reference to a basis of counterterms involving traces and de Donder tensors  $\cD := \prd \vf - \tfrac{1}{2} \vf^{\, \pe}$. 
\begin{table}[h]
\label{table:1}
\centering
\begin{tabular}{c|cccc}
\diagbox{$\vf^{\, \pe}$}{$\cD$} & 0 & 1 & 2 & 3 \\
\hline
 0 & $ \vf \ \, \vf \ \, \vf $ &  $\cD \ \vf \ \vf$ & $\cD \, \cD \ \vf$ & $\cD \, \cD \, \cD$ \\ 
 1 & $\vf^{\, \pe}\, \vf \ \vf$ & $\vf^{\, \pe}\, \cD \ \vf$ & $\cD \, \cD \, \vf^{\, \pe}$ & \\
 2 & $\vf^{\, \pe}\, \vf^{\, \pe} \, \vf$ & $\vf^{\, \pe} \, \vf^{\, \pe} \, \cD$ & &\\
 3 & $\vf^{\, \pe}\, \vf^{\, \pe}\, \vf^{\, \pe} $ & & & \\
\end{tabular}
\caption{Building blocks of  Fronsdal cubic vertices}
\end{table}

In the Fronsdal setting, in particular, typically all types of terms collected in Table \ref{table:1} actually enter the cubic vertex. (See {\it e.g.} \cite{Manvelyan:2010je,Mkrtchyan:2011uh}.) For Maxwell-like Lagrangians, differently, due to the simplicity of the free kinetic term \eqref{Ml}, {\it one never needs to introduce traces in the procedure}. Indeed, in the reducible framework that is of most interest for us, traces essentially represent independent fields that may or may not be considered in the construction. In a minimal scheme one can avoid introducing them altogether, reducing the types of terms to be taken into account just to the first row of Table \ref{table:1}, as summarised in Table \ref{2}, with the ``de Donder tensors'' $\cD$ here to be identified with divergences of the fields, $ \cD \, := \, \prd \vf \, $.
\begin{table}
\centering
\begin{tabular}{c|cccc} 
\diagbox{$\vf^{\, \pe}$}{$\cD$} & 0 & 1 & 2 & 3 \\
\hline
 0 & $ \vf \, \vf \,\vf $ &  $\cD\, \vf \, \vf$ & $\cD \, \cD \, \vf$ & $\cD \, \cD \, \cD$ \\ 
\end{tabular}
\caption{Building blocks of Maxwell-like cubic vertices} \label{2}
\end{table}
\\
Since at the TT-level there is no difference between our construction and the corresponding analysis for Fronsdal fields, these vertices are found to admit a number of derivatives bound to satisfy the usual double inequality \cite{Bengtsson:1986kh, Metsaev:2005ar, Manvelyan:2010jr}
\be \label{bound}
s_1 \, + \, s_2 \, + \, s_3 \, - \, 2 \mbox{min}\{s_1,\, s_2,\, s_3 \} \, \leq \, \# \, \pr \, \leq \,  + \, s_1 \, + \, s_2 \, + \, s_3 \, .
\ee
However, one has to remind that each Maxwell-like vertex provides a synthetic description of several cross-interactions involving low-spin particles. For the latter in general the total number of derivatives would exceed the bound, thus implying that in the full theory one should deal with the issue of clarifying the role of all those additional couplings. Cubic couplings for reducible systems were investigated from a different perspective in \cite{Bengtsson:1987jt, Buchbinder:2006eq, Fotopoulos:2007nm, Fotopoulos:2007yq}.

\subsection*{Adapted Noether procedure and deformation of the constraints} \label{sec: cubic_gen}

The second feature that we would like to stress concerns the need for deforming the transversality condition \eqref{transversality}.

Once all possible counterterms encoded in Table \ref{2} are considered, one needs to identify the deformation of the free gauge symmetry accounting for cubic-level gauge invariance. The possibility to drive the procedure to completion is tied to the form of the variation of the resulting cubic Lagrangian $\cL_1$, that can be schematically written as follows
\be \label{dL1}
\d \, \cL_1 \, \sim \,  \D_1 \, M \, + \, \D_2 \, \prd \prd \vf\, ,
\ee
where $\D_1$ and $\D_2$ are local operators depending on fields, gauge parameters and derivatives, whose form we compute explicitly. The peculiar aspects of our procedure are encoded in the last term in \eqref{dL1}, proportional to the double divergence of the field: although vanishing on the free mass shell, it cannot be  {\it locally} related to the free equations of motion, and thus cannot be absorbed in the contribution to \eqref{dL1} proportional to $M$. For instance, for spin $s = 2$,
\be
\prd \prd M \, = \, - \, \Box \, \prd \prd \vf \, ,
\ee
and similarly for higher spins, with higher divergences of $M$ to be involved. For this reason, following the steps of the usual Noether procedure, it is no longer clear, and in general it won't be true, that one can obtain from \eqref{dL1} the correction to the gauge transformation $\d_1 \vf$ in its standard local form.  On the other hand, the variation of the free Lagrangian 
\be \label{deltaL0}
\d \, \cL_0 \, \sim \prd \e \, \prd \prd \vf \, ,
\ee
shows that the contributions in $\D_2$ can be compensated by a suitable deformation of the transversality constraint \eqref{transversality} of the form
\be \label{Constraint deformation}
\prd \e \, + \, \cO \, (\vf, \, \e) \, = \, 0 \, ,
\ee
where in $\cO \, (\vf, \, \e)$ all fields and gauge parameters may enter, {\it a priori}.
For the spin$-2$ case, where \eqref{Ml} provides the linearised equations of traceful unimodular gravity \cite{Alvarez:2006uu}, the corrections \eqref{Constraint deformation} are instrumental to reproduce the covariant form of the transversality condition, $\cD \cdot \e = 0$, and thus to ultimately recover the underlying geometry. It may be interesting to notice that, in this special case, the Noether procedure at cubic order would actually {\it not} compel the introduction of corrections to \eqref{transversality}. This shows, in our opinion, that including in the Noether procedure the possibility encoded in \eqref{Constraint deformation} in principle retains a deeper meaning than just allowing to enforce some algebraic cancellations.

For higher spins, \eqref{Constraint deformation} provides an additional equation that enters the perturbative reconstruction of the gauge structure and that turns out to be necessary to the completion of the procedure. We judge that this option may be of more general interest. In this spirit, we shall present a systematic discussion of how to include in the Noether procedure perturbative corrections to possible constraints to be imposed on the free gauge symmetry.

\subsection*{Partially reducible theories} \label{sec: partred}

The Maxwell-like equations \eqref{Ml} propagate the maximal reducible unitary representation of the Lorentz group encoded in the symmetric tensor $\vf$, and one may wonder whether consistent truncations of the spectrum may be implemented.  The degrees of freedom of the various particles propagating in \eqref{Ml} are essentially contained in the traces of the field $\vf$.\ft{An off-shell covariant separation of the various single-particle components would combine traces with multiple divergences. See Section $4$ of \cite{ML} for a detailed discussion of this point.} Thus, a natural guess is that partial truncations of the spectrum may be implemented by trace constraints of increasing strength. In this view we suggest that the projected equations of motion
\be
M_k \, = \, M \, + \, \l_k \, \h^k \, \prd \prd \vf^{\, [k -1]} \,  = \, 0 \, ,
\ee
where $\vf^{\, [m]}$ denotes th $m-$th trace of $\vf$ while $\l_k$ is a coefficient ensuring that $M_k$ be $k-$traceless, should describe particles with spin $s, s-2, \ldots,$ up to $s - 2(k - 1)$, provided that, on top of the transversality condition \eqref{transversality}, the $k-$th traces of the field and of the gauge parameter be vanishing:
\be \label{tracek}
\vf^{\, [k]} \, = \, 0\, , \hskip 2cm  \e^{\, [k]} \, = \, 0 \, .
\ee
In this view, the fully reducible theory described by \eqref{Ml} and its fully irreducible counterpart, obtained upon imposing tracelessness of both $\vf$ and $\e$ \cite{SV}, would represent just the extrema of a chain of theories describing spectra of decreasing complexity. Up to relatively simple modifications, our construction of the cubic vertices apply to each of these options.

Let us mention that models where each tensor carries the degrees of freedom of a number of particles that increases with its rank provide instances of higher-spin theories for which a full non-linear counterpart is not known. Indeed, in any putative Maxwell-like complete theory, that contains at least one copy of a Maxwell-like field for each (even) rank, there would appear  infinitely many massless particles for any given value of the spin, up to those that may be eliminated from the spectrum by conditions of the form \eqref{tracek}. This is to be contrasted with the presently known Vasiliev's theories whose spectra involve at most a finite number of particles with the same spin \cite{Vasiliev:1990en,Vasiliev:2003ev, bciv, Didenko:2014dwa}.

\subsection*{Plan of the paper} \label{sec: plan}

In Section \ref{sec: maxwell} we review the Maxwell-like theory of \cite{ML} and present its partially reducible generalisations. In Section \ref{sec: noether} we rephrase the Noether procedure in a way that allows to encompass the case of constrained gauge symmetries in a systematic fashion. The computation of the cubic vertex is presented in Section \ref{sec: cubic}. In particular, in Section \ref{sec: cubic_TT} we discuss the TT sector, while Section \ref{sec: cubic_red} contains a detailed illustration of the subsequent steps for the fully reducible case. In section \ref{sec: cubic_irred} we  discuss how to adapt the construction of vertices to the full class of theories with partially reducible spectra, up to the fully irreducible one. We shall comment on the possibility of including traces in the vertices in Section \ref{sec: cubic_traces}.  Section \ref{sec: AdS} is devoted to the construction of (A)dS vertices. Here we switch to a different language, that of ambient space, that has the advantage of allowing to effectively bypass the need for computing commutators of covariant derivatives, at the price of making the whole construction less explicit. In the Outlook we collect our final comments, while four appendices contain technical remarks on our notation, on the absence of alternatives to the deformation of the constraint in our context, and on the spectra of a few selected partially reducible models.


\section{Maxwell-like description of higher spins} \label{sec: maxwell}

In this section we illustrate the formulation of Maxwell-like higher-spin theories. We provide a review of the fully reducible model of \cite{ML} together with a proposal for partially reducible Maxwell-like theories, covering all possible unitary spectra encoded in principle in a symmetric tensor.

The covariant description of massless representations of the Poincar\'e group with a given finite spin (helicity) $s$ is encoded in the Fierz system \cite{Fierz},
\begin{align} \label{fierz}
&\Box  \, \vf \, = \, 0\, ,& & & &\Box  \, \e\, = \, 0\, \, ,\nonumber\\
&\prd \vf \, = \,  0 \, ,& &\vf \, \sim \, \vf + \pr \, \e& &\prd \e\, = \,  0 \, , \\
&\vf^{\, \pe} \,   = \, 0 \, , & & &
&\e^{\, \pe} \,   = \, 0 \,  , \nonumber
\end{align}
and one can look for its possible off-shell completions under the requirements of locality, gauge invariance and, for simplicity, absence of auxiliary fields. Once the  d'Alembert equations in \eqref{fierz} are relaxed to  $\Box \vf \neq 0$ and $\Box \e\neq 0$ then $\prd \vf$ is no longer gauge invariant and one is forced to consider configurations such that  $\prd \vf \neq 0$. Compensating the gauge variation of $\Box \vf$ requires indeed to combine it with the divergence of $\vf$ to form the tensor
\be \label{M}
M \, = \, \Box \, \vf \, - \, \pr \, \prd \vf\, , 
\ee
that in this sense provides the minimal building block of any gauge theory. Its gauge variation is
\be \label{deltaM}
\d \, M \, = \, - \, 2 \, \pr^{\, 2}\, \prd \e\, . 
\ee
The remaining conditions in \eqref{fierz} may or may not be relaxed and the various options lead to theories with different particle contents and possibly different Lagrangian formulations. As far as the structure of the kinetic tensor is concerned the simplest choice is to fully relax tracelessness of $\vf$ and $\e$ while keeping the transversality condition on the latter, and thus consider the gauge-invariant Lagrangian with constrained gauge symmetry 
\begin{align} 
&\cL_0 \, = \, \12 \, \vf \, ( \Box \, - \, \pr \, \prd)\, \vf \, ,   \label{ML} \\
&\prd \e\, = \, 0 \, .   \label{tdiff}
\end{align}
The corresponding equations of motion, $M = 0$,  propagate a reducible spectrum of massless particles with spin $s \, - \, 2k$, $k = 0, \, \ldots,\, [\tfrac{s}{2}]$.

We put forward that the same Lagrangian \eqref{ML} can be used to describe spectra of various degree of complexity  if, on top of the transversality condition  \eqref{tdiff}, one also requires both the gauge field and the gauge parameter to be subject to trace conditions of varying strength:
\be \label{highk}
\begin{split}
&\vf^{\, [k]} \, = \, 0,  \hskip 2cm \e^{\, [k]} \, = \, 0 \, , \\
\end{split}
\ee
where $k$ can take any of the values between $k = 1$ and $k = [\tfrac{s}{2}]$.  In particular, $k = 1$ provides the strongest possible trace constraint that one can impose. In the latter case the theory can be interpreted as an allowed gauge fixing of the Fronsdal Lagrangian and correspondingly the spectrum is known to collapse to the single massless particle of spin $s$ \cite{SV}. 

On the basis of the fact that the physical polarisations of the fully reducible system are carried on-shell by the traces of $\vf$  \cite{ML}, we argue that higher values of $k$ in \eqref{highk} should correspond to less severe truncations of the spectrum. In this view, for instance, the system of \eqref{ML}, \eqref{tdiff} and \eqref{highk} for $k = 2$  should describe a pair of massless particles with spin $s$ and $s-2$, respectively. For  $k=3$ one additional particle of spin $s-4$ is expected to enter the spectrum and so forth, while the weakest trace condition, corresponding to $k = [\tfrac{s}{2}]$, should lead to a reducible theory where only the lowest-spin particle, a scalar or a vector, gets eliminated from the spectrum.

It is to be stressed that the Lagrangian always retains the same form \eqref{ML}, irrespective of the strength of the trace condition imposed in  \eqref{highk}, while the equations of motion resulting after implementing \eqref{highk} obtain as the $k-$th traceless projection of \eqref{M} and look
\be \label{eomirr}
M_{\, k} \, = \, M \,  + \, \fr{2\, k}{\prod_{i=1}^k [D + 2(s - k -i)]} \, \h^{\, k} \, \prd \prd \vf^{\, [k-1]} \, = \, 0 \, .
\ee
In Appendix \ref{D} we analyse in detail the spectra of a few non-trivial cases, providing quantitative support to our intuition. Let us also mention that for spin $s = 2$ \eqref{ML} and \eqref{tdiff} encode the linearised version of unimodular gravity, with or without the additional scalar mode depending on whether the trace of the graviton is kept or discarded. (See {\it e.g.} \cite{Alvarez:2006uu} and references therein.)

The unique local alternative to imposing the transversality condition \eqref{tdiff} is to include the trace of $\vf$ in the kinetic operator, thus leading to  the Fronsdal tensor 
\be \label{F}
\cF \, = \, \Box \, \vf \, - \, \pr \, \prd \vf\, + \, \pr^{\, 2} \, \vf^{\, \pe}\, ,
\ee
whose gauge invariance holds if 
\be \label{trace}
\e^{\, \pe} = 0\, . 
\ee
The equations $\cF = 0$ propagate a single massless particle of spin $s$, and can be derived from a Lagrangian assuming the additional constraint $\vf^{\, \pe \pe} = 0$ \cite{fronsdal}. Both the transversality condition \eqref{tdiff} and the trace constraint \eqref{trace} can be easily evaded by introducing additional, Stueckelberg compensator fields $D$ and $\a$ that transform as follows
\begin{align}
& \d \, D \, = \, \prd \e\, , \\
& \d \, \a \, \, = \, \e^{\, \pe} \, ,
\end{align}
so as to guarantee unconstrained gauge invariance of the corresponding extended kinetic tensors 
\begin{align}
& M  \quad \longrightarrow  \quad M \, + \, 2 \, \pr^{\, 2} \, D \, ,\label{tripletM} \\
& \cF \,  \quad \longrightarrow  \quad \cF \, - \, 3 \, \pr^{\, 3} \, \a \, .\label{minimal} 
\end{align}

In the case of \eqref{minimal}, upon introducing an additional auxiliary field $\b$ serving as a Lagrange multiplier for the double trace of $\vf$, one recovers the minimal unconstrained formulation of \cite{fs3, fms1}. Other unconstrained extensions of the Fronsdal theory with bigger field content (and only two-derivative kinetic tensors) were proposed in \cite{pt, bpt, quartet, dario07}. The unconstrained completion of $M$ provided by \eqref{tripletM}, on the other hand, may be further supplemented with an additional auxiliary field $C$ transforming as $\d C = \Box \e$, so as to recover the triplet Lagrangian emerging from free tensionless strings, as first shown in \cite{triplet1, triplet2} and further elaborated upon in \cite{triplet3, fs2, st, ft}, whose spectrum corresponds to that of  the fully reducible Maxwell-like theory with no trace conditions imposed. The tensors \eqref{M} and \eqref{F} somehow provide higher-spin counterparts of the kinetic tensors of  the Maxwell and of the (linearised) Einstein theories, respectively. The idea that the spin-one and the spin-two models may have different higher-spin incarnations was first put forward in \cite{D10, dariokyoto} and further elaborated upon in \cite{Dconnections,Joung:2012qy, DXN}. Reducible higher-spin models were also given a free frame-like formulation on flat and (A)dS backgrounds in \cite{Sorokin:2008tf}, while related constructions stemming from a  worldline perspective were presented in \cite{Bastianelli:2012nh, Bastianelli:2015tha}.

Our goal is to study the interactions among fields whose free Lagrangian is \eqref{ML}. Since the full effect of possible trace constraints would be to give rise to projected equations of motion while leaving the free Lagrangian untouched, the structure of the corresponding cubic vertex turns out to be universal for the whole class of partially reducible models, with differences emerging only at the level of their gauge deformations. 

 From our perspective one relevant issue concerns the role of the transversality condition \eqref{tdiff} in the deformation procedure. As we shall see in Section {\ref{sec: cubic}, in general we are able to find a local solution for the cubic vertices  only upon allowing for corrections to  \eqref{tdiff}. (See also Appendix \ref{C}.) Thus, in order to better frame the details of our computations, in the next section we illustrate the general framework for adapting the Noether procedure to the case of constrained gauge symmetries.

\section{On the Noether procedure for constrained gauge theories} \label{sec: noether}

\subsection{Generalities} \label{sec: gen}

 The basic assumption  underlying the Noether procedure\ft{The Noether procedure has been widely used in recent years in the higher-spin literature. Here we quote the work that, to our knowledge, first implemented it to the purpose of investigating interactions among higher-spin  massless fields.} \cite{Berends:1984rq} is that both the action functional $S [\vf]$ and its gauge invariances $\d \, \vf$  admit a perturbative expansion in powers of the fields:
\be 
\begin{split}
& S \, [\vf] \, = \, S_0 \, [\vf] \, + \, g\, S_1 \, [\vf] \, + \, g^{\,2}\, S_2 \, [\vf] \, \ldots \, ,\\ 
& \d \, \vf \, = \, \d_0 \, \vf \, + \, g \, \d_1 \, \vf  \, + \, g^{\, 2} \, \d_2 \, \vf  \, \ldots \, ,
\end{split}
\ee
where in this section $\vf$ (and later $\e$) represents a collective symbol denoting all types of fields (and parameters) entering the theory. The coupling $g$ plays the role of a counting parameter for the powers of fields to be added to the corresponding zero-th order quantity. $S_0 \, [\vf]$ is the free action depending quadratically on the fields, while $S_k\, [\vf]$ involves $k$ additional powers of the fields: $S_k \, [\vf] \sim \vf^{\, k + 2}$. Similarly, $\d_0 \, \vf$ represents the Abelian gauge symmetry of the free action, in our case $\d_0 \, \vf = \pr \, \e$,  while the higher order contributions to the transformations are linear in the gauge parameter (since in the present context we always consider infinitesimal gauge transformations) and depend in principle on an increasing number of fields: $\d \vf_k  \sim \, \e \, \vf^{\, k}$.

 Asking for perturbative gauge invariance of the action  leads to the Noether system of equations,
\be \label{noether}
\begin{split}
& \d_0 \, S_0 \, [\vf] \, = \, 0 \, ,\\ 
& \d_1 \, S_0 \, [\vf] \, + \, \d_0 \, S_1 \, [\vf] \, = \, 0\, ,\\
& \d_2 \, S_0 \, [\vf] \, + \, \d_1 \, S_1 \, [\vf] \, + \, \d_0 \, S_2 \, [\vf] \,= \, 0\, ,\\
& \dots \, 
\end{split}
\ee
whose solutions provide in principle the possible interaction vertices compatible with gauge invariance, while also determining the corresponding deformations, Abelian or non-Abelian, of the free gauge symmetry. The procedure just outlined applies to the case of unconstrained gauge symmetries or even when constraints are present that, anyway, do not get deformed themselves.

 On general grounds, however, one may consider the possibility that the free gauge parameters are subject to given off-shell conditions implemented through the action of some (linear) operator ${\cal O}$ in the schematic form
\be \label{constraints}
{\cal O} \e \, = \, 0,
\ee
concrete examples being provided by the transversality and trace conditions \eqref{tdiff} and  \eqref{trace}, respectively.  What we would like to stress is that the constraints \eqref{constraints}  may  themselves receive perturbative corrections in increasing powers of the fields:
\be \label{corr}
{\cal O} \e \, + \, g \, {\cal O}_1 \, (\e\, ,\vf) \, + \, g^2 \, {\cal O}_2 (\e\, , \vf^{\, 2}) \, + \, \ldots\, = \, 0 \, .
\ee
In a minimal scheme where one does not consider the option of trading \eqref{constraints} for the inclusion of auxiliary fields, in order to properly take the corrections encoded in \eqref{corr}  into account one has to modify the Noether equations \eqref{noether}.  

In particular, from this perspective, the terms encoding the deformation of the gauge symmetry order by order, denoted with $\d_k \, \vf$, would admit themselves a perturbative expansion in powers of $\vf$, due to their {\it implicit} dependence on $\vf$ hidden in $\e$ because of \eqref{corr}. In order to display the effective dependence of the various terms by powers of the fields we shall make use of the following notation:
\be
\begin{split}
&\d_k \, \vf \, = \, \d_k^{\, (0)} \vf \, + \,  \d_k^{\, (1)} \vf \, + \, \d_k^{\, (2)} \, \vf \, + \, \ldots \, , \\
& \d_k^{\, (l)} \vf \, := O\, (\vf^{\, k+l}) \,
\begin{cases}
&\mbox{{\it explicitly} on} \sim \vf^{\, k},    \\
&\mbox{{\it implicitly} on} \sim \vf^{\, l} ,\, \mbox{via its $\e-$dependence.}
\end{cases}
\end{split}
\ee
Correspondingly, the terms entering the Noether system \eqref{noether} get modified as follows
\begin{align} \label{Noether}
& o\, (\e, \vf):  \d \, {\cal S} \, = \, \int \, \frac{\d {\cal L}_0}{\d \vf} \,  \d_0^{\, (0)} \vf, \nonumber \\ 
& o\, (\e, \vf^{\, 2}):   \d \, {\cal S} \, = \, \int  \, \left\{\frac{\d {\cal L}_0}{\d \vf} \,(\d_0^{\, (1)} \vf \, + \, \d_1^{\, (0)} \vf) \, + \,
\frac{\d {\cal L}_1}{\d \vf} \,  \d_0^{\, (0)} \vf \right\}, \\
& o\, (\e, \vf^{\, 3}):   \d \, {\cal S} \, = \, \int \, \left\{\frac{\d {\cal L}_0}{\d \vf} \,(\d_0^{\, (2)} \vf  +  \d_1^{\, (1)} \vf +  \d_2^{\, (0)} \vf ) + 
\frac{\d {\cal L}_1}{\d \vf}  (\d_0^{\, (1)} \vf  +  \d_1^{\, (0)} \vf ) +  \frac{\d {\cal L}_2}{\d \vf}  \d_0^{\, (0)} \vf \right\} \, \nonumber ,\\
& \dots \, . \nonumber
\end{align}
For instance, in the case under scrutiny in this paper, both $\d_0^{\, (0)} \vf$ and  $\d_0^{\, (1)} \vf$ are given by the Abelian transformation  $\pr \, \e$. On the other hand, while the parameter $\e$ in $\d_0^{\, (0)} \vf$ solves $\prd \e = 0$, it contributes to $\d_0^{\, (1)} \vf$, and thus to the second equation in \eqref{Noether}, only via the solution to
\be \label{corr1}
\prd \e\, + \, g \, {\cal O}_1 \, (\e\, ,\vf) = 0\, ,
\ee
to first order in $\vf$, in case a non-trivial correction term ${\cal O}_1 \, (\e\, ,\vf)$ is considered. Similarly for the higher-order terms. Let us recall that here $\e$ denotes collectively {\it all} the gauge parameters entering the procedure and in general the correction terms  $\cO_k \, (\e, \, \vf^k)$ may depend linearly on all of them. 

As usual, from \eqref{Noether} one can first determine the cubic vertices $\cL_1$ on the free mass shell, {\it i.e.} for those configurations that vanish when $ \tfrac{\d {\cal L}_0}{\d \vf}  =  0$ holds. At this stage the corrections encoded in \eqref{corr} do not play any special role. Once $\cL_1$ is found one should proceed to collect all terms in $\d \cS$ that are quadratic in $\vf$  and that vanish on the free mass shell, and arrange them as in the second equation of \eqref{Noether} so as to compute $\d_1 \vf$. The latter in general, as one can see in \eqref{Noether}, comprises two contributions, whose splitting may well be not uniquely determined. It seems to us that two situations are possible:
\begin{description}
 \item[(1)] in some cases the nature of the constraints \eqref{constraints} may be such that it is always possible to reabsorb any correction encoded in \eqref{corr} in a suitable {\it local} deformation of $\d \, \vf$ while keeping \eqref{constraints} unchanged;
 \item[(2)] in other cases corrections of the form \eqref{corr} and correspondingly modified Noether equations  \eqref{Noether} may be unavoidable in order to grant for the existence of a {\it local } solution to the deformation procedure.  
\end{description} 
If option {\bf (1)} occurs, then the general system \eqref{Noether} encoding the corrections \eqref{corr} can be equivalently traded for the more customary Noether system \eqref{noether}, with constraints on the gauge parameters kept in their original ``free'' form \eqref{constraints}. However, we would like to stress that even in these cases taking the deformations into account can be relevant. They may provide some algebraic simplifications, to begin with, but more important than that, as suggested by the perturbative reconstruction of unimodular gravity (see Section \ref{sec: cubic}), they may be connected with the underlying geometry of the theory. On the other hand, whenever {\bf (2)} holds, addressing the procedure in its generalised form \eqref{Noether} becomes unescapable, as far as one wishes to keep manifest locality at the perturbative level. 
  
While the procedure just outlined applies in principle to all sort of constraints possibly imposed on gauge parameters off shell, in the following we shall comment more in detail on the two cases that are directly relevant for higher spins, from our perspective.

\subsection{Trace constraints} \label{sec: trace}

Algebraic constraints, like the trace conditions \eqref{trace} of the Fronsdal theory, are usually regarded  as harmless from the perspective of the deformation procedure, since on Minkowski background there appears to be no need to deform them in order to get the corresponding covariant form of the first-order deformation \cite{Boulanger:2006gr}, \cite{Manvelyan:2010jr, Sagnotti:2010at}. From our perspective, this feature of the Fronsdal theory may be envisaged considering the variation of the Fronsdal Lagrangian for spin $s = 3$,
\be \label{s3}
\d \, \cL \, = \, - \, \frac{3}{2} \, \e^{\, \pe} \, \prd \, \cF^{\, \pe} \, ,
\ee
from which it is manifest that deformations of the trace conditions may only affect terms that are {\it locally} proportional to the free equations of motion and that, as such, can be equivalently included in a local redefinition of $\d_1^{\, (0)} \vf$. For these reasons, it seems plausible to us that, more generally, the option {\bf (1)} may more easily refer to constraints \eqref{constraints} that are algebraic in nature.  
 
However, a few remarks are in order:
\begin{itemize} 
\item to begin with, it has to be stressed that the cubic deformation of the spin--three Fronsdal theory on (A)dS background computed in \cite{Zinoviev:2008ck} was indeed found to require a deformation of the trace constraint on $\e_{\, \m \n}$, taking into account perturbative corrections to the background metric tensor, used to compute the trace, via additional dynamical spin--two contributions. From this perspective, similar considerations should apply to the double-trace conditions to be enforced on the Fronsdal  gauge fields off shell.\ft{It is to be mentioned that this option does not manifest itself in the frame-like formulation, where trace conditions are encoded in the choice of the frame fields and of the generalised spin connections as taking values in irreps of the orthogonal group in the tangent space. (See {\it e.g.} \cite{bciv, Didenko:2014dwa}.)}
\item Similar metric-deformations of the trace constraint were also considered in a three-dimensional setup  in \cite{Campoleoni:2012hp}, for the case of a Fronsdal spin--three field coupled to gravity, motivated by the need to maintain manifest diffeomorphism invariance. In \cite{Fredenhagen:2014oua} and  \cite{Campoleoni:2014tfa} the analysis was repeated from a frame-like perspective and extended to include spin--four cubic self-interactions together with $3-3-4$ vertices. Interestingly, there it was found that deriving the metric-like vertices from the frame formulation (where interactions are known in closed form in $D = 3$) naturally leads to deformations of the trace constraints also involving higher-spin fields. 
\item Indeed, it is only on (A)dS backgrounds that cubic couplings of the form $2-s-s$ entail a {\it minimal coupling} vertex. In this sense (at least for a covariant and local theory) only in (A)dS can one interpret the gauge deformation on the spin--two side really as the first non-linear contribution to full diffeomorphism invariance. In the flat case, on the other hand, the allowed cubic couplings are non-minimal, and thus one should not necessarily expect to be forced to treat $h_{\, \m \n}$ as a fluctuation of the metric tensor from the very beginning. In other words, it may not be obvious {\it a priori} that the Minkowski metric $\h_{\, \m \n}$ be corrected by powers of the rank--two tensor $h_{\, \m \n}$.
\end{itemize}
These general considerations notwithstanding, we think that there are reasons to expect the general status of trace conditions to be more complicated, both for (A)dS and for Minkowski backgrounds.

To begin with, in the same sense as the deformation of the Fronsdal trace constraint found in (A)dS could be expected on account of the existence of spin--two fluctuations, higher-spin covariance of the full theory leads us to expect that in general the full deformation of the trace conditions should involve also higher spins and not just be determined by a metric correction to the trace, consistently with the three-dimensional findings of \cite{Campoleoni:2014tfa}. From our vantage point, as summarised in our comment to \eqref{s3}, the expectation is that those deformations may well be reabsorbed by a local field redefinition (at the cubic level at least this option should manifest itself both on flat and on (A)dS backgrounds), but it is quite tantalising that they naturally emerge in $D = 3$  when deducing the metric-like theory from its complete frame-like counterpart.

For similar reasons, one may wonder whether the absence of deformation for the Fronsdal trace constraint on Minkowski background may be a spurious effect of performing the analysis only up to the cubic level. Indeed, from the perspective of the  unconstrained theory of \cite{fs3}, absence of deformation for the condition $\e^{\, \pe} = 0$ would be tantamount to absence of Noether corrections for the gauge transformation of the compensator field $\a$ there introduced (see \eqref{minimal}). Although not impossible in principle we have no reason to expect it in general. 

In addition, the case of spin $s \geq 4$ appears to be more involved than the example of spin $3$ that we referred to in \eqref{s3}, due to the interplay between presence or absence of the double trace constraint and actual form of the corresponding free equations of motion.

Finally, as already mentioned  (see also Section \ref{sec: gauge5}), in the context of unimodular gravity knowledge of the full non-linear theory explicitly indicates that the transversality constraint on the parameter has to be eventually covariantised,
\be
\prd \e\, = \, 0 \quad \longrightarrow \quad \cD \cdot \e\, = \, 0 \, ,
\ee
although at cubic level in Minkowski space one finds that this option is not forced upon by consistency of the Noether procedure.

\subsection{Divergence constraints} \label{sec: divergence}

As we shall see in the next section, option {\bf (2)} is realised by the Maxwell-like systems described by \eqref{ML} and \eqref{tdiff}. Since we are going to discuss this point in detail, here we shall limit ourselves to make just a few remarks.

We could not find in general a local solution to the deformation procedure without taking the corrections to the constraints \eqref{tdiff} into account.  As we discuss in Appendix \ref{C}, our conclusion is that this feature is intrinsic to the systems under consideration: starting with the free Maxwell-like Lagrangians \eqref{ML} it is unavoidable to implement the deformation of the transversality condition \eqref{tdiff} in order to grant for local cubic-level gauge invariance. The main underlying reason is that, for Maxwell-like systems, there are field configurations that vanish on the free mass shell in force of conditions that are {\it non--locally} proportional to the equations of motion. These aspects will be also discussed in Section \ref{sec: cubic}.

It should be mentioned, however, that an alternative option for Maxwell-like theories would be to {\it solve} for the transversality condition \eqref{tdiff} in terms of unconstrained gauge parameters, to then proceed with the implementation of the Nother method in the standard fashion. The general solution to \eqref{tdiff} was computed in \cite{FLS} and takes the form
\be \label{solve0}
\e_{\, \m_1 \, \cdots \m_{s-1}} \, = \, \pr^{\, \a_1} \, \cdots \, \pr^{\, \a_{s-1}}\, \e^{\, (0)}_{\, \a_1 \, \cdots \a_{s-1}, \, \m_1 \, \cdots \m_{s-1}}\, ,
\ee
with $\e^{(0)}$ taking values in the irrep of $GL(D)$ corresponding to a rectangular tableau with two rows,
\be \label{epsilon}
\e^{(0)}_{\, \a_1 \, \cdots \, \a_{s-1}, \, \m_1 \, \cdots\, \m_{s-1}}:\, \hskip .2cm {\small
\overbrace{\young(\hfil \hfil \cdots \hfil,\hfil \hfil \cdots \hfil)}^{s-1} } \, 
\ee
In this parametrisation the gauge symmetry is reducible, the pattern of gauge-for-gauge transformations being encoded in the following chain of  reducibility conditions
\be \label{gfg}
\d \, \e^{\, (k)}_{\, \b^{(k)}, \, \cdots,  \, \b^{(1)}, \,  \a_{s-1}, \,  \m_{s - 1}} \, = \,
\pr^{\, \b^{(k+1)}} \, \e^{\, (k+1)}_{\, \b^{(k+1)}, \, \b^{(k)}, \, \cdots,  \, \b^{(1)}, \,  \a_{s-1}, \,  \m_{s - 1}}\, ,
\ee
with $k = 0, \, \cdots, \, D-2$, where the symmetries of the parameters $ \e^{\, (k)}$ are encoded in the $GL(D)-$tableaux\ft{Actually, as observed in \cite{FLS}, it is always possible to parametrise the gauge symmetry of a theory (at the linear level, at least) in infinitely many different, yet equivalent ways. Even for ordinary, spin$-1$ Maxwell fields one may write the scalar gauge parameter as a contracted, multiple divergence of a symmetric tensor of arbitrary rank, $ \d \, A_{\, \m} \, = \, \partial_{\, \m} \, \partial^{\, \a_1} \, \cdots \, \partial^{\, \a_s} \, \e_{\, \a_1 \, \ldots \, \a_s}$, obtaining in this way a rather unusual-looking high-derivative and reducible parametrization of the gauge symmetry of a free massless vector field. For the Maxwell-like case this observation also shows that, contrarily to standard lore, a fully unconstrained, local description of massless metric-like higher spins is possible without invoking auxiliary fields in the construction.}
\be \label{epsilonk} 
\e^{\, (k)}_{\, \b^{(k)}, \, \cdots,  \, \b^{(1)}, \,  \a_{s-1}, \,  \m_{s - 1}}:\, \hskip .3cm
{\small \overbrace{\young(\hfil \hfil \cdots \hfil,\hfil \hfil \cdots \hfil,1,\vdots,k)}^{s-1}} \,  \\
\ee
However, the customary system of Noether equations \eqref{noether} applied to this high-derivative and reducible gauge symmetry would eventually result to be far more complicated to deal with than the equivalent one \eqref{Noether}, where the constraint is not solved for and its corrections are taken into account. 

Moreover, while the two options for $\e_{\, \m_1 \, \ldots \, \m_s}$, either satisfying \eqref{tdiff} or being solved for as in \eqref{solve0}, are certainly locally equivalent, it may expected in general that they may differ when {\it global} issues are considered. 

In addition, we see no reasons in principle why it should be always possible to solve any constrained gauge symmetry in terms of {\it local} unconstrained  alternative parametrisations. In the absence of such solutions, once again, it seems to us that the only options would be either to look for suitable sets of auxiliary fields entering the theory as substitutes of the condition \eqref{constraints}, or to implement the Noether procedure as encoded in  \eqref{Noether}. 
  
\section{The cubic vertex } \label{sec: cubic}

In this section we construct the cubic vertex for Maxwell-like fields in flat space-time. Our goal is to solve the Noether system \eqref{Noether} to first-order in the parameter $g$, which requires to compute three types of terms:
\begin{itemize}
 \item the cubic vertex $\cL_1$;
 \item the explicit deformation of the gauge transformation $\d^{\, (0)}_1  \vf$;
 \item the first correction to the transversality constraint: $\prd \e\, + \, g \, \cO_1 \, (\e, \vf) \, = \, 0$.
\end{itemize} 

\subsection{TT-sector} \label{sec: cubic_TT}

In this section we briefly recall how to determine the general form of the transverse-traceless part of the cubic vertex. At this level there are no differences with the case of Fronsdal fields and for this reason we shall only sketch the general structure of the computation. For more details see \cite{Manvelyan:2010jr}, whose conventions we follow here, together with Appendices \ref{A} and \ref{B}. We aim to construct the cubic interactions among three gauge fields of arbitrary spins,
\be
\vf_1 \, := \, \vf^{(s_1)}\, (x_1; a)\, , \hskip 1cm \vf_2 \, := \, \vf^{(s_2)}\, (x_2; b)\, , \hskip 1cm \vf_3 \, := \, \vf^{(s_3)}\, (x_3; c) \, ,
\ee
starting from the so-called TT sector of the vertex, including no divergences nor traces, that we write in the schematic form
\be \label{TT}
\cL^{\, TT}_1 \, = \, \sum_n \, K_{\{Q, \, n\}} \, \int \, d^D \, \m \, T\,(Q, \, n)\, \vf_1 \, \vf_2 \, \vf_3 \, ,
\ee
where 
\begin{itemize}
 \item $T\,(Q, \, n) \, := \, T\,(Q_{12}\, , Q_{23}\, , \, Q_{31}; \, n_1, \, n_2, \, n_3) $ 
 
is the operator performing all the contractions of indices and containing the derivatives assumed to enter the vertex. Its general form is
\be \label{T}
T\,(Q, \, n)\, = \, (\pr_a \cdot \pr_2)^{\, n_1} \, (\pr_b \cdot \pr_3)^{\, n_2} \, (\pr_c \cdot \pr_1)^{\, n_3} \, (\pr_a \cdot \pr_b)^{\, Q_{12}} \, (\pr_b \cdot \pr_c)^{\, Q_{23}} \,  (\pr_c \cdot \pr_a)^{\, Q_{31}}\, ,
\ee  
where $n$ is the total number of derivatives, while the various coefficients are linked by the following relations
\be \label{coefficients}
\begin{split}
& n_1 + \, n_2 + \, n_3 \, = \, n \, ,                      \\
& Q_{12} \, + \, Q_{31} \, + \, n_1 \, = \, s_1 \, , \\
& Q_{12} \, + \, Q_{23} \, + \, n_2 \, = \, s_2  \, , \\
& Q_{23} \, + \, Q_{31} \, + \, n_3 \, = \, s_3  \, .
\end{split}
\ee
Vertices that differ by total derivatives are to be regarded as equivalent, and it is convenient to establish a convention allowing to systematically deal  with integration by parts. Our definition \eqref{T} for $T\,(Q, \, n)$ corresponds to the {\it cyclic ansatz}, defined such that a derivative operator gets always contracted with the field that precedes it cyclically, {\it i.e.} fields and derivatives always appear in the following order:
\be
\vf^{\, (s_i)} \, (x_i) \, \cdot \fr{\pr}{\pr x^{i + 1}}\, .
\ee
The space-time arguments of each field are taken to be different in order to simplify the various manipulations. 
 \item $d^D \, \m \, := \, d^D \, x_2\, d^D \, x_3 \, \d^D (x_1 - x_2) \,  \d^D (x_2 - x_3) $

is the integration measure, containing the delta-functions that eventually compute the various fields at coinciding points, thus  ensuring locality of the result.
 \item $K_{\, \{Q, n\}} \, := \, K_{\, \{Q_{12}\, , Q_{23}\, , \, Q_{31}; \, n_1, \, n_2, \, n_3\}}$ 
 
are the relative coefficients among the various terms of the vertex and represents the unknowns of our initial problem. Due to the relations \eqref{coefficients} they effectively depend only on the $Q_{ij}$'s. (Or on the $n_i$'s.) Correspondingly, the sum in \eqref{TT} effectively runs only over one set of independent options.
\end{itemize}

Let us observe that in the cyclic ansatz d'Alembertian operators acting on fields are automatically excluded, consistently with the fact that, at cubic level, they can always be reabsorbed (up to terms proportional to traces and divergences of the fields) by a field redefinition of the free Lagrangian of the schematic form $\vf \, \to \, \vf + \, \vf \, \vf$.\ft{This observation alone explains the upper bound on the total number of derivatives that can appear in the cubic vertex:
$$
n \, \leq \, s_1 + \, s_2 + \, s_3, 
$$
since any number of derivatives higher than the sum of the spins would generate d'Alembertian operators, upon integration by parts.  One should also keep in mind, however, that field redefinitions will propagate their effects to higher orders, and it may be not obvious {\it a priori} that a simplification at the cubic level would correspond to an overall simplification of the full theory. All in all, making a choice about field redefinitions amounts to choosing a basis of fields, and the general convenience of a given option may be not at all apparent at the cubic level, or at any given finite order.}

The coefficients $K_{\, \{Q, n\}}$ can be fixed, up to an overall constant, by requiring that the variation of  \eqref{TT} under the gauge transformation of each field, 
\be \label{deltaphi}
\d \, \vf^{\, (s)}\, (x; \, a)\, = \, s \, a \cdot \pr \, \e^{\, (s - 1)} \, (x; \, a) 
\ee
vanishes, up to 
\begin{itemize}
 \item total derivatives,
 \item contributions that vanish when the free equations of motion hold,
 \item terms containing traces or divergences of the fields, as well as contributions proportional either to $\Box \e$ or to $\prd \e$, whose cancellation requires the introduction of counterterms proportional to traces or divergences of the fields.
\end{itemize} 
The result is 
\be
\begin{split}
K_{\, \{Q, n\}} \, = \, \fr{k}{Q_{12} !\, Q_{23} !\,Q_{31} !}\, ,
\label{Q}
\end{split}
\ee
where the overall constant $k$ is expected to be fixed by the next order in the Noether procedure (see \cite{Metsaev:1991mt, Metsaev:1991nb} for earlier results in the light-cone formulation, and  \cite{Sleight:2016dba} for a related discussion from a holographic perspective.)

A simple consequence of \eqref{Q}, as briefly explained in \cite{Manvelyan:2010jr}, is that the number of derivatives $n$ in the vertex cannot be lower than $s_1+s_2+s_3-2 \min\{s_1,s_2,s_3\}$. To see this we first recall that the sum of the $Q$'s is fixed by the spins of the fields involved in the vertex and by the number of derivatives, since \eqref{coefficients} implies
\be
s_1+s_2+s_3=n+2(Q_{12}+Q_{23}+Q_{31})\,.\label{sQ}
\ee
 Now, from \eqref{Q} we deduce that, for gauge invariance to hold,  the sum in \eqref{TT} has to run over {\it all non-negative values of the $Q$'s, satisfying \eqref{coefficients}}.  In particular, there are non-vanishing coefficients in \eqref{Q} with one of the $Q$'s equal to zero. Let us take for example $Q_{12}=0$. Obviously, $Q_{23}+Q_{31}\leq s_3$, from which it follows, that $Q_{12}+Q_{23}+Q_{31}\leq s_3$ has to hold for all values of the $Q$'s, otherwise one of the terms in the ansatz \eqref{TT} will have negative powers of contractions. Similarly, one can show that $Q_{12}+Q_{23}+Q_{31}\leq s_i$ for any $i=1,2,3$. This in turn implies that
\be
n=s_1+s_2+s_3-2(Q_{12}+Q_{23}+Q_{31})\geq s_1+s_2+s_3-2 \min\{s_1,s_2,s_3\}\, , \label{LowerBound}
\ee
consistently with the light-cone analysis of \cite{Metsaev:1991mt, Metsaev:1991nb}.

The ensuing steps in the calculation depend on the gauge variation of the TT part of the vertex, that we write schematically as follows:
\be \label{delta1}
\begin{split}
& \d \, \cL^{TT}_1  = \,  \sum_Q \,  K_{\, \{Q, n\}} \, \int d^D \, \m \\
&\biggl\{\fr{s_1 n_1}{2} \, T(n_1 - 1) \, (\Box_3 - \Box_2 - \Box_1) \, \e_1 \, \vf_2 \, \vf_3 \, - \,s_1 \,s_2 \,Q_{12} \, T(Q_{12} - 1) \, \e_1 \, \cD_2 \, \vf_3 \, + \biggr. \\
&\ \ \, \fr{s_2 n_2}{2} \, T(n_2 - 1) \, (\Box_1 - \Box_2 - \Box_3) \, \vf_1 \, \e_2 \, \vf_3  \, - \, s_2 \,s_3 \,Q_{23} \, T(Q_{23} - 1) \, \vf_1 \, \e_2 \, \cD_3 \, +  \\
&\ \  \biggl. \fr{s_3 n_3}{2} \, T(n_3 - 1) \, (\Box_2 - \Box_3 - \Box_1) \, \vf_1 \, \vf_2 \, \e_3  \, - \, s_3 \,s_1 \,Q_{31} \, T(Q_{31} - 1) \, \cD_1 \, \vf_2 \, \e_3 \biggr\} \, , 
\end{split}
\ee
where in particular all arguments in $T\,(Q, \, n)$ that are not explicitly displayed are meant to be unchanged. 

Our goal is to compensate \eqref{delta1} by means of suitable counterterms and, if needed, by also reconsidering the transversality condition \eqref{tdiff}. We shall illustrate the corresponding results in the next two sections.
 
\subsection{Completion of the vertex} \label{sec: cubic_red}

The full cubic vertex will turn out to have the following schematic form,
\be \label{LDDD2}
\cL_1 \, = \, \cL^{\, TT}_{1} \, + \,  \cL_{1, \cD} \, + \,  \cL_{1, \cD \cD} \, + \,  \cL_{1, \cD \cD \cD} \, ,
\ee
where in particular the various counterterms will involve up to three divergences (here denoted with $\cD$) but no traces of $\vf$.   The gauge variation of the TT part of the vertex, computed in \eqref{delta1}, suggests a systematic way of introducing these counterterms. Here we shall illustrate the main steps of the computation, also addressing the reader to Appendix \ref{B} for further technical details.
\subsubsection*{$\cL_{\, 1,\,  \cD}$} \label{sec: L1D}
In order to select possible counterterms to be added to \eqref{TT} we focus first on the terms in \eqref{delta1} containing d'Alembertian operators acting on gauge parameters. (Those proportional to $\sim \Box \vf$ can be traded for free equations of motion and enter the deformation of the gauge symmetry.). Given that at the free level
\be
\d \, \cD \, = \, \Box \, \e \, ,
\ee
we see that terms of the type $\sim \Box \e\, \vf\, \vf$ are to be compensated by monomials of the form $\sim \cD \, \vf \,\vf$. The explicit computation fixes the corresponding coefficients  as follows:
\be \label{orderD}
\begin{split}
\cL_{1,\, \cD} \, = \, \sum_Q \,  K_{\, \{Q, n\}} \, & \int d^D \, \m \, \left\{ \fr{s_1 \,n_1}{2} \, T(n_{2} - 1) \, \cD_1 \, \vf_2 \, \vf_3 \, + \right. \\
& \, \left.\fr{s_2 \,n_2}{2} \, T(n_{3} - 1) \, \vf_1 \, \cD_2 \, \vf_3 \, +
\, \fr{s_3 \,n_3}{2} \, T(n_{1} - 1) \, \vf_1 \, \vf_2 \, \cD_3 \right\} \, .
\end{split}
\ee
Clearly, the cancellation mechanism has to work in the same fashion for the three fields (whose ranks may well be not all distinct), so that in the following we can limit ourselves to illustrating the procedure with reference to one of them, say $\vf_1$. 

The gauge variation of  $\cL^{\, TT}_{1} \, + \, \cL_{1,\, \cD}$ with respect to free Abelian transformation of $\vf_1$, that we denote with  $\d_0 \vf_1 := \, \d^{\vf_1}_{\, 0}$, is
\begingroup
\allowdisplaybreaks
\begin{align} \label{delta2}
& \d^{\vf_1}_{\, 0} \, \biggl\{\cL^{\, TT}_{1} \, + \, \cL_{1,\, \cD}\biggr\} \, = \,  \int d^D \, \m \left\{ \sum_n \,  K_{\, \{Q, n\}}   \biggl(\fr{s_1 s_2 n_1 n_2}{2} \, T\,(n_1 - 1,\, n_2 - 1)  \, \e_1 \, \cD_2 \, \Box \, \vf_3 \biggr. \right. \nonumber \\
& \biggl. - \, \fr{s_1 s_2 n_1 n_2}{2} \, T\, (n_1 - 1,\, n_2 - 1) \, \Box \, \e_1 \, \cD_2 \, \vf_3 \, - \, \fr{s_1 s_3 n_3 n_1}{2} \, T\, (n_3 - 1,\, n_1 - 1)  \, \Box \, \e_1 \, \vf_2 \, \cD_3 \biggr) \nonumber \\
& + \, \sum_n \,  \fr{s_1 s_2 s_3}{2} \, \biggl(K_{\, n_1 - 1} \, n_1 \,  T\, (n_1 - 1,\, Q_{23} - 1) \, - \, K_{\, n_3 - 1} \, n_3 \,  T\, (n_3 - 1,\, Q_{21} - 1)  \biggr) \, 
\e_1 \, \cD_2 \, \cD_3 \nonumber \\
&- \, \left.\sum_n \,  \fr{s_1 s_2 (s_2 - 1)}{2} \, K_{\, n_3 - 1} \, n_2 \,  T\, (n_2 - 1,\, Q_{12} - 1) \, \e_1 \, \prd \cD_2 \, \vf_3 \, \right\}\, .
\end{align}
\endgroup
\subsubsection*{$\cL_{\, 1,\,  \cD \cD}$} \label{sec: L1DD}

From the second line of \eqref{delta2} , in particular, one can read the structure of the counterterms containing $\cD_1 \cD_2 \vf_3$ and $\cD_1 \vf_2 \cD_3$
needed for the second step. Taking the variations with respect to $\vf_2$ and $\vf_3$ into account  one finds eventually
\be \label{orderDD}
\begin{split}
\cL_{1,\, \cD \cD} \, = \, \sum_n \,  K_{\, \{Q, n\}} \, \int d^D \, \m \, & \left\{\fr{s_1 s_2 n_1 n_2}{2} \, T(n_1 - 1, \, n_2 - 1) \, \cD_1 \, \cD_2 \, \vf_3  \right. \\
&\, \, \, + \fr{s_1 s_3 n_3 n_1}{2} \, T(n_3 - 1, \, n_1 - 1) \, \cD_1 \, \vf_2 \, \cD_3 \\
&\, \, \, + \left. \fr{s_1 s_3 n_3 n_2}{2} \, T(n_3 - 1, \, n_2 - 1) \, \vf_1 \, \cD_2 \, \cD_3 \right\} \, .
\end{split}
\ee
The overall gauge variation of the cubic vertex to this order contains, besides terms proportional to the free equations of motion, the following terms:
\begin{align} \label{deltaDD}
&\d^{\vf_1}_{\, 0} \, \biggl\{\cL_{1, 0} \, + \, \cL_{1,\, \cD} \, + \, \cL_{1,\, \cD \cD}\biggr\}\, = \nonumber \\
& = \,  \int d^D \, \m \left\{ - \sum_n \,  K_{\, \{Q, n\}} \, \fr{s_1 s_2 s_3 n_3 n_1 n_2}{2} \, T\,(n_3 - 1,\, n_1 - 1,\, n_2 - 1)  \, \Box \e_1 \, \cD_2 \, \cD_3  
\right.\nonumber \\
&  \left. - \sum_n \,  K_{\, \{Q, n\}} \, \fr{s_1 s_2 s_3}{2}\, n_3 \, n_2\, Q_{12}\, T\,(n_3 - 1,\, n_2 - 1, \, Q_{12} - 1)\, (s_2 - 1)\, \e_1 \, \prd \cD_2 \, \vf_3 \right. 
\nonumber \\
&  \left. - \sum_n \,  K_{\, \{Q, n\}} \, \fr{s_1 s_2 (s_2 - 1)}{2}\, n_2 \, T\,(n_3 - 1, \, Q_{12} - 1)  \, \e_1 \, \prd \cD_2 \, \vf_3 \right\} \, .
\end{align}
%

\subsubsection*{$\cL_{\, 1, \,  \cD \cD \cD}$} \label{sec: L1DDD}

In order to compensate the contribution containing $\sim \Box \e \, \cD \, \cD$ in \eqref{deltaDD} we introduce the last term  in our chain of compensations 
\be \label{orderDDD}
\cL_{1,\, \cD \cD \cD} \, = \, \sum_n \,  K_{\, \{Q, n\}} \, \int d^D \, \m \, \fr{s_1 s_2 s_3 n_3 n_2 n_1}{2} \, T(n_3 - 1, \, n_2 - 1, \, n_2 - 1) \, \cD_1 \, \cD_2 \, \cD_3 \, ,
\ee
thus completing the construction of the Maxwell-like cubic vertex  \eqref{LDDD2},  via \eqref{TT}, \eqref{Q}, \eqref{orderD}, \eqref{orderDD} and \eqref{orderDDD}.

Before moving to analysing the gauge structure of the theory, let us pause to comment on our result in comparison with related literature. 

Cubic interactions for reducible higher-spin systems have been investigated using BRST techniques in \cite{Buchbinder:2006eq,Fotopoulos:2007nm,Fotopoulos:2008ka,Fotopoulos:2010ay}, in the so-called triplet formulation \cite{triplet1,triplet2,triplet3,fs2,st,ft} related to the free tensionless limit of string theory and involving  additional, unphysical fields. Our focus is on reducible models involving physical tensors only, those containing the propagating polarisations of each particle. 

Our main motivation is essentially one of simplicity, in the following sense: focusing on physical tensors we manage to keep the resulting action as simple as possible, since already at the free level the additional fields of the triplet description make the theory more involved, the difference being quite significant if one compares the corresponding free Lagrangians for mixed-symmetry fields \cite{ML, st}. It was not guaranteed by any general arguments that the simplicity of the free action would be kept at the cubic level, but our results  confirm it. The absence of auxiliary fields implies that from our vertices we can better read the physical couplings, while at the technical level it dispenses us with the need to compute the deformation of the gauge symmetry for a wider class of tensors. Moreover, the fully reducible Maxwell-like theory appears to admit an easy generalisation to a whole hierarchy of reducible theories with spectra of decreasing complexity, as we illustrate in Section \ref{sec: cubic_irred}. All of these theories are described by the same free Lagrangian and correspondingly deform with the same cubic vertices, the differences among them being visible only at the level of the gauge deformation, where the structure of the free equations of motion becomes relevant.

The choice of working with the minimal Lagrangian \eqref{ML}, however,  brings in a novelty with respect to the standard implementation of the Noether procedure. As anticipated in the Section \ref{sec: noether}, compensating the variation of \eqref{LDDD2}  will require to rethink the conditions that make the free Lagrangian gauge invariant, and specifically to introduce field-dependent corrections to the transversality condition \eqref{tdiff}.  

\subsection{Conditions for gauge invariance} \label{sec: gauge3}

Let us compute once more the variation of  \eqref{LDDD2}, now collecting all terms that we could not compensate otherwise. In order to properly discuss and appreciate the outcome at this level we consider the {\it full} variation of $\cL_1$, under the simultaneous transformation of the three fields. Schematically,
\be \label{deltaL}
\d \, \cL_1 \, = \, \sum_n \,  K_{\, \{Q, n\}} \, \int d^D \, \m \sum_{i = 1,2,3} \, \biggl\{\D^{(i)}_1 \, M_i \, + \, \D^{(i)}_2 \, \prd \cD_i \biggr\}
\ee
where $\D^{(i)}_1$ and $\D^{(i)}_2$ are operators depending on all the fields and all the parameters other than $\vf_i$ and $\e_i$, whose explicit form is the following: 
\be
\begin{split}
\D^{(1)}_1 \, M_1 \, = \, &+ \,  \frac{s_2 \, n_2}{2}\, T\, (n_2-1) \, M_1 \, \e_2 \, \vf_3 \, -\, \frac{s_3 \, n_3}{2}\, T(n_3 -1)\, M_1\, \vf_2 \, \e_3  \\
& + \, \frac{n_2 \, n_3 \, s_2 \, s_3}{2}\, T\, (n_2-1,\, n_3-1)\, M_1\, \e_2\, \cD_3\, , \label{DM}
\end{split}
\ee
\be
\begin{split}
\D^{(1)}_2 \, \prd \cD_1 \, = & - \, n_1\, s_1 \, (s_1 -1) \,  s_3 \,Q_{31}  \, T(n_1-1, \, Q_{31}-1) \, \prd \cD_1 \, \vf_2 \, \e_3  \\
& - \, n_1\, n_2\, (s_1-1) \,  \frac{s_1 \, s_2 \, s_3}{2}\,\, Q_{31}  \,  T(n_1 - 1, \, n_2-1, \, Q_{31}-1)\, \prd \cD_1 \, \cD_2 \, \e_3 \, , \label{DD}
\end{split}
\ee
and similarly, by cyclicity, for $i = 2, 3$. As anticipated, the two terms in \eqref{deltaL} play crucially different roles:
\begin{itemize}
\item On the one hand, they both correctly vanish on the free mass-shell, since in particular one can show that \cite{ML}
\be
M \, = \, 0 \, \qquad \Rightarrow \qquad \prd \cD \, := \, \prd \prd \vf \, = \, 0 \, .
\ee
\item According to our adapted Noether equations \eqref{Noether}, from \eqref{DM} we can compute the contribution to the first-order deformation of the type $\d_1^{\, (0)} \vf$, performing suitable integration by parts:
\begin{align} 
& \D^{(1)}_1 \, M_1 \, = \, - \, M_1 \, (a) \, *_a \, \d^{(0)}_1\, \vf^{(s_1)}\, (a) \, , \label{Mdelta} \\
& \d^{\, (0)}_{1} \, \vf^{\, (s_{1})} \, = \, -\,  \sum_{n_i} \, K_{\, {n_i}} \, \int d^D \, \m \, s_1! \, \left\{
\fr{s_2 \, n_2}{2} \, T_a \, (n_2 - 1) \, \e_2 \, \vf_3 \right. \nonumber \\
&\left.  - \, \fr{s_3 \, n_3}{2} \, T_a \, (n_3 - 1) \, \vf_2 \, \e_3 \, + \, \fr{n_2 \, n_3 \, s_2 \, s_3}{2} \, T_a \, (n_2 - 1, \, n_3 - 1) \, \e_2 \, \cD_3 \, \right\}\, , \label{deltaphizero}   
\end{align}
where the notation $T_a \, (Q_{\, ij}\, , n_k)$, with the subscript $a$, stands for the operator that one gets from \eqref{T} upon performing the substitution
$ \pr_{\, a} \, \longrightarrow \, a$, and where the derivatives $\pr_1$ with respect to the first argument are understood to be integrated by parts. \end{itemize}
In the standard Noether procedure \eqref{Mdelta} would provide the only contribution to $\d \cL_1$; the presence of the additional terms \eqref{DD} represent the main novelty of our analysis. We compute their form in the next section.

\subsection{Deformation of the transversality constraint} \label{sec: gauge4}

The main observation concerning the role of the terms \eqref{DD} in the variation of the cubic Lagrangian is that {\it the double divergences of the various fields, in general, can be expressed in terms of the corresponding tensors $M$ only non-locally.} This is essentially due to the Bianchi identity for the Maxwell-like tensor $M$ that relates it to a symmetrised gradient of the double divergence of $\vf$,
\be \label{bianchi}
\prd M \, = \,- \,  \pr \, \prd \prd \vf \, ,
\ee
as a manifestation of the transverse gauge invariance of the corresponding Lagrangian. For this reason if we were to follow the standard Noether prescription in this setting and look for a solution of the same form as \eqref{Mdelta} above, 
\be
\D^{(1)}_2 \, \prd \cD_1 \, = \, M_1 \, (a) \, *_a \, \tilde{\d}^{(0)}_1\, \vf^{(s_1)}\, (a) \,
\ee
we would find that  the contribution $\tilde{\d}^{(0)}_1\, \vf^{(s_1)}\, $ to the deformation of the gauge transformation would be non-local. This is the reason why, in order to achieve gauge-invariance of the cubic vertex in a fully local setting, we have to resort to the possibility that the transversality constraint gets itself corrected by field-dependent terms. 

To begin with, for the sake of completeness, let us write the full set of these terms, generated by the total variation of the cubic Lagrangian:
\be \label{DDfull}
\begin{split}
\d_0\, \cL_1\, & =\, O\, (M) \, + \, \sum_{n_i}^{}\,  \int d^D \m \,K_{(n_i)}\left\{\right.\\
&\left.\left[ -s_{1}Q_{12}s_{2}(s_{2}-1)n_{2}T(n_{2}-1|Q_{12}-1)\e^{(s_{1}-1)}\prd \cD^{(s_{2}-2)}\vf^{(s_{3})}+ \right.\right. \\
&\left.\,\,\left. -s_{2}Q_{23}s_{3}(s_{3}-1)n_{3}T(n_{3}-1|Q_{23}-1)\vf^{(s_{1})}\e^{(s_{2}-1)}\prd \cD^{(s_{3}-2)}+ \right.\right. \\
&\left.\,\,\left. -s_{3}Q_{31}s_{1}(s_{1}-1)n_{1}T(n_{1}-1|Q_{31}-1)\prd \cD^{(s_{1}-2)}\vf^{(s_{2})}\e^{(s_\cD^{3}-1)} \right]\right.+\\
&\left.\,\,\,+\dfrac{s_{1}s_{2}s_{3}}{2}\left[ -Q_{12}(s_{2}-1)n_{2}n_{3}T(n_{2}-1,n_{3}-1|Q_{12}-1)\e^{(s_{1}-1)}\prd \cD^{(s_{2}-2)}\cD^{(s_{3}-1)}+\right.\right.\\
&\,\,\q\q\q\q\,\,\left.\left. -Q_{23}(s_{3}-1)n_{3}n_{1}T(n_{1}-1,n_{3}-1|Q_{23}-1)\cD^{(s_{1}-1)}\e^{(s_{2}-1)}\prd \cD^{(s_{3}-2)}+\right.\right.\\
&\,\,\q\q\q\q\,\,\left.\left. -Q_{31}(s_{1}-1)n_{1}n_{2}T(n_{1}-1,n_{2}-1|Q_{31}-1)\prd \cD^{(s_{1}-2)}\cD^{(s_{2}-1)}\e^{(s_{3}-1)}\,\,\right]\right\}\, .
\end{split}
\ee
Integrating by parts, we can always write \eqref{DDfull} in a form where the contributions in  $\prd \cD$ are factorised, {\it e.g.} for \eqref{DD}
\be
\D^{(1)}_2 \, \prd \cD_1 \, = \, \prd \cD_1 \, (a) \, *_a \, \tilde{\d}^{(1)}_0\, \vf^{(s_1)}\, (a) \,,
\ee
so as to make it manifest the correction to \eqref{tdiff}, by comparison with  the gauge transformation of the free Lagrangian
\be \label{deltafree}
\d \, \cL_0 \, = \, - \, s \, \left({s \atop 2}\right) \, \prd \e\, \prd \cD \, .
\ee
Let us go back, for simplicity, to the terms containing the double divergences of the first field\footnote{In our cyclic ansatz, they emerge when we variate the Lagrangian with respect to the third field.}: inverting the differential operator $T$ we can write this contribution in the following schematic way:
\be
\label{rimanenzaschematica}
\d_0^{\, \vf_1}\, \cL_1\, =\, \prd\cD_1\,\cC_{1}\, (\vf_2,\, \e_3)\,,
\ee
where $\cC_1$ is linear in both the gauge parameter and the field. Performing all integrations by parts and comparing with \eqref{deltafree} we eventually find 
\be \label{deformation}
\begin{split}
\prd \e_1 \, &= \, g \, \frac{1}{s_1 \, (s_1 - 1)} \, \cC_{1}\, (\vf_{2},\, \e_3) \, \\
& = \, -g\,\sum_{n}\int d^D \m \,K_{(n_i)}s_{3}Q_{31} \, (s_1 - 2)!\,n_{1} \\
&\left[+T_a(n_{1}-1|Q_{31}-1)\vf^{(s_{2})}\e^{(s_{3}-1)}+\frac{n_{2}s_{2}}{2}T_a(n_{1}-1,n_{2}-1|Q_{31}-1)\cD^{(s_{2}-1)}\e^{(s_{3}-1)}\right]\,,
\end{split}
\ee
where  the operator $T_a$ is to be understood as in \eqref{deltaphizero}. 

One may observe that additional fields may turn the adapted Noether procedure into a conventional one, as for the case of the triplet construction of \cite{Fotopoulos:2010ay}. However, as already mentioned, trading the issue of deforming the constraint for that of computing the gauge deformation of additional fields is not obviously more convenient in general. Moreover, in order to precisely assess  the relation between the two descriptions one should be either integrate or gauge away the unphysical fields from the cubic vertices of the triplets, an issue which appears more involved than performing the same task at the free level.

To reiterate, the deformation of the transversality condition is only necessary whenever it is not possible to reconstruct in the corresponding terms tensors that are locally proportional to the equations of motion. Thus, for instance, whenever one is able to reproduce the symmetrised gradient of a double divergence, then in force of \eqref{bianchi} it is possible to make the tensor $M$ to appear in a local expression and thus the corresponding contribution may be included in $\d_1^{\, (0)}$. Moreover, the explicit form of the deformation, of course, will depend on the ansatz chosen for the cubic vertex (the cyclic one, in our case). However, as we show in Appendix \ref{C}, regardless of the initial ansatz, it is not possible in general to eliminate the need for the deformation altogether.

\subsection{Additional couplings and field redefinitions} \label{sec: gauge5}

Having at one's disposal the option to deform the transversality constraint \eqref{tdiff} makes it apparently possible to enforce cubic-level gauge invariance for an additional class of vertices, that do not descend from the chain of compensations stemming from the TT part of the cubic Lagrangian.  Indeed, one may consider cubic couplings of the following type
\be \label{L*sss}
\cL^{\, *}_{s_1-s_2-s_3} \, = \, \l \,  \prd \prd \vf_{s_1} \, \vf_{\, s_2} \, \vf_{\, s_3} \, ,
\ee
whenever the various indices allow to form a scalar. Exploiting the variation of the free Lagrangian for the first field \eqref{deltaL0}, cubic level gauge invariance obtains upon enforcing the following condition
\be \label{dsss}
\prd \e_{\, s_1}\, - \, \fr{\l}{s_1 \, (s_1 - 1)} \, \left(\pr \, \e_{\, s_2} \,\vf_{\, s_3} \, +  \,\vf_{\, s_2} \, \pr \,\e_{\, s_3}  \right)\, = \, 0 \, .
\ee
Other examples involving more derivatives can also be constructed along the same lines. 

The main feature of this class of couplings that we would like to stress is that, in general, \ft{In a number of cases vertices of this type directly qualify as due to local field redefinitions.  Writing the vertex in the form $\cL^{\, *}_{s_1-s_2-s_3}  \sim  \cO \left(\vf_{\, s_1}  \vf_{\, s_2}  \vf_{\, s_3} \, \right)$, where the operator $\cO$ in particular generates the double divergence of at least one field $\vf_{\, s_k}$, the redefinitions of the free theory can be identified as those for which the operator $\cO$ admits the following solution
\be \label{tildeO}
\cO \, = \, \left[\tilde{\cO}, \, a_i  \cdot \pr_{\, i} \, (\pr_i \, \cdot \pr_{\, a_i})^{\, 2} \right] \, ,
\ee
for at least one of the fields, where $i =1, 2, 3$  (no summation implied over $i$). The operator $a_i  \cdot \pr_{\, i} \, (\pr_i \, \cdot \pr_{\, a_i})^{\, 2}$ acting on $\vf_i$ generates the divergence of the free equations of motion. The most general operator $\cO$ generating fake interactions may involve redefinitions of more than one of the fields involved,  that are understood to be included in $\tilde{\cO}$.} they can be interpreted as due to {\it non-local} field redefinitions of the free theory, in force of the relation between the double (or multiple) divergence of a field and its free equations of motion stemming from the ``Bianchi identity'' of the Maxwell-like tensor
\be \label{dM}
\prd M \, = \, - \, \pr \, \prd \prd \vf \, .
\ee
In this sense, it looks like they may have some physical relevance, in spite of the number of derivatives being below the bound \eqref{bound}.\ft{Two-derivative vertices involving massless, symmetric higher spins were identified in the light-cone analysis of \cite{Bengtsson:1986kh, Metsaev:1991mt,Metsaev:1991nb, Bengtsson:2014qza}. However, on the one hand their existence is specific of $D=4$, while the possibility to put them in covariant form by conventional methods appears to be problematic \cite{Conde:2016izb,Sleight:2016xqq, Ponomarev:2016lrm}.} For instance, one could make use of this observation to construct a quartic-complete Abelian, spin$-2$ Lagrangian (higher-spin generalizations are immediate) of the form
\be \label{quartic1}
\cL \, = \, \12 \, h_{\, \m \n} \, M^{\, \m \n} \, + \, g \, \prd \prd h \, h_{\, \m \n} \, h^{\, \m \n} \, - \, \12 \, g^{\, 2} \, h_{\, \r \s} \, h^{\, \r \s} \, 
\Box \, (h_{\, \m \n} \, h^{\, \m \n}) \, ,
\ee
whose gauge invariance at cubic level holds upon enforcing the following deformation of the transversality constraint
\be
\prd \e \, = \, g \, (\pr_{\, \m} \e_{\, \n} \, + \, \pr_{\, \n} \, \e_{\, \m}) \, h^{\, \m \n}\, .
\ee
The interactions terms in \eqref{quartic1} can indeed be generated by the following non-local redefinition of variables in the free Lagrangian:
\be \label{nonloc}
h_{\, \m \n} \ \longrightarrow \ h_{\, \m \n} \, + \, g \, \fr{\pr_{\, \m}\, \pr_{\, \n}}{\Box} \, (h_{\, \r \s} \, h^{\, \r \s})\, .
\ee
However, the construction of the corresponding quartic theory in the unconstrained setting of \cite{fs2, st, ft} indicates that the vertices actually arise after a local field redefinition of the auxiliary field present in that context. For the spin--two case, for instance, the free Lagrangian can be written
\be \label{triplet}
\cL \, (\vf, D) \, = \, \12 \, h_{\, \m \n} \, M^{\, \m \n} \, + \, 2 \, \prd \prd h \, D \, - \, 2 \, D \, \Box \, D\, ,
\ee
where gauge invariance holds under $\d h_{\, \m \n} = \pr_{\, (\m} \e_{\, \n)}$ and $\d D  = \prd \e$. It is then possible to recognise that the vertices in \eqref{quartic1} emerge from \eqref{triplet} after field-redefining $D$ and its gauge transformation as follows
\be
\begin{split}
& D \ \longrightarrow \ D \, + \, \12 \, g \, h_{\, \m \n} \, h^{\, \m \n}  \, ,\\
& \d \, D \, = \, \prd \e \, - \, g \, \pr_{\, (\m} \e_{\, \n)} \, h^{\, \m \n}\, ,
\end{split}
\ee
and then performing the non-linear gauge fixing $D = 0$. The most plausible conclusion stemming from these observations is that, although non-locally generated by \eqref{nonloc}, the vertices of \eqref{quartic1} should not really correspond to something non trivial in the cohomology of the corresponding kinetic operator. From a different perspective  one may observe that, in a theory with constrained gauge symmetry, some non-local manipulations may be allowed without altering the physical properties of the theory, whenever they have a counterpart in local manipulations on pure-gauge auxiliary fields appearing with a non-trivial kinetic operator \cite{D10, ontherel}.

 Let us also mention, however, that the case of self-interacting spin$-2$ fields leads to consider that the two-derivative vertices of the type \eqref{L*sss}, or more generally the full mechanism of deformation of the transversality constraint \eqref{tdiff}, may retain a deeper meaning than just being a substitute for a missing Stueckelberg field. On the one hand one may observe that, in this case, such couplings do  arise as part of the chain of compensations stemming from the transverse-traceless vertex. Indeed,  proceeding as we discussed in Section \ref{sec: cubic} one finds the following term 
\be
\d L^{\, TT}_{2-2-2} \,  \sim \, \prd \prd h \, h_{\, \m \n} \, \pr^{\, \m} \, \e^{\, \n} \, ,
\ee
and the standard way to compensate it, in the spirit of the ordinary Noether procedure, would amount  to just complete the vertex adding the following contribution:
\be \label{L*2}
\sim \, \prd \prd h \, h_{\, \m \n} \, h^{\, \m \n} \, .
\ee
On the other hand, from our present vantage point, we would like to stress  that an equivalent result would obtain by enforcing the deformation of the constraint,
\be \label{d*2}
\prd \e \, + \, k \, \, h_{\, \m \n} \, \pr^{\, \m} \, \e^{\, \n} \,  = \, 0 \, ,
\ee
for a suitable choice of $k$. The main observation in this respect is that, although not needed in order to grant for cubic-level gauge invariance,  the deformation \eqref{d*2} is actually crucial to obtain the covariantised constraint, 
\be
\prd \e \, = \, 0 \, \hskip 1cm \longrightarrow \hskip 1cm \cD \cdot \e \, = \,  0
\ee
and thus to the recovery of the underlying geometry, in particular through the reconstruction of the (inverse) metric: $\h^{\, \m \n} \, \pr_{\, \m} \, \e_{\, \n} \, \longrightarrow \, g^{\, \m \n} \pr_{\, \m} \e_{\, \n}$.

What this example suggests, in our opinion, is that deforming the constraint may have to do with basic features of the theory even when not forced by consistency requirements, such as locality. After all, a field redefinition amounts to a change of basis in the space of fields, and simplicity of the perturbative construction may turn out to be a deceptive criterion on which to found the general procedure, possibly hiding more important properties that may become visible only at higher orders, if not just in the full theory.

\section{Partially reducible and irreducible cases} \label{sec: cubic_irred}

In this section we discuss the construction of cubic vertices for the whole set of partially reducible models presented in Section \ref{sec: maxwell}.

 The main observation is that, since traces do not enter the Noether procedure that we implemented in Section \ref{sec: cubic}, one can simply reiterate it in this context and construct in the same fashion the cubic interactions for all these partially reducible theories. The only modification to be taken into account is the different form of the equations of motion for $k-$traceless tensors, that we report here,
\be \label{eomirr2}
M_{\, k} \, = \, M \,  + \, \fr{2\, k}{\prod_{i=1}^k [D + 2(s - k -i)]} \, \h^{\, k} \, \prd \prd \vf^{\, [k-1]} \, = \, 0 \, ,
\ee
implying the appearance of new terms proportional to traces of the double divergence of the field. Schematically, in the fully reducible case, after putting together all terms proportional to the equations of motion, we obtain
\be
\d_0 \, \cL_1 \, = \, -\, \d_1\vf*M \, + \, \prd\prd \vf*\cC_1 \, ,
\ee
where we used the $*-$product defined in Appendix \ref{A} to perform the contractions and where $\cC_1$ is the first deformation of the transversality constraint. 

The main observation is that we can make use of the same result for the class of partially reducible theories, upon substituting $M$ with  \eqref{eomirr} as follows
\be
\d_1\vf*M \, \q\rightarrow\q \,  \d_1\vf*M_{\, k} \, \, -\, \d_1\vf* \fr{2\, k}{\prod_{i=1}^k [D + 2(s - k -i)]} \, \h^{\, k} \, \prd \prd \vf^{\, [k-1]} \, .
\ee
This substitution produces an additional term contributing to the deformation of the constraint. Eventually, keeping in mind that $\prd\prd\vf^{\, [k-1]} $ and $M_{\, k}$ are $k-$traceless we get:
\be
\begin{split}
& (\d_1\vf)_{\, k}\, =\, \cT_k \, \d_1\vf \, \\
&\cC_1^{k} \, =\, \cT_k \, \left\{ \cC_1+\dfrac{s(s-1)}{\prod_{i=1}^k [D + 2(s - k -i)]} \, \h^{\, k-1} \, (\d_1\vf)^{\, [k]}\right\}\, ,
\end{split}
\ee
where $\cT_k$ is the projector to the $k-$th traceless part of a rank$-s$ tensor.\ft{The projector to the traceless part was computed, for instance, in \cite{fms1}. Higher trace projections can be obtained in similar fashion.} Thus, the only differences emerging from choosing different trace conditions in \eqref{highk} reside in the structure of the gauge deformation and do not touch the form of the vertex \eqref{LDDD2}.

\section{Inclusion of traces} \label{sec: cubic_traces}


So far we were mainly concerned with the construction of ``minimal'' interactions, where the traces of each tensor entering the cubic vertex were not taken into account. The issue is relevant since the traces of the various fields are related to the lower-spin particles present in the (partially) reducible spectrum of each free Lagrangian. In this section we want to discuss how to systematically include vertices containing traces of the fields. We shall refer to the fully reducible Maxwell-like theory, but it should be understood that similar considerations apply to the partially reducible theories discussed in Section \ref{sec: maxwell} and in Section \ref{sec: cubic_irred}.

The free gauge transformation of the $k-$th trace of a Maxwell-like rank$-s$ tensor $\vf$ is
\be
\d \, \vf^{\, [k]} \, = \, \pr \, \e^{\, [k]} \, ,
\ee
with $\e^{\, [k[}$ such that
\be
\prd \e^{\, [k[} \, = \, 0\, .
\ee
Moreover, due to the fact the $M = 0 \, \longrightarrow\, \prd \prd \vf = 0$, it is easy to recognise that $\vf^{\, [k]}$ satisfies
\be
\Box \, \vf^{\, [k]} \, - \, \pr \, \prd \vf^{\, [k]} \, = \, 0 \, .
\ee
It is then manifest that $\vf^{\, [k]}$ behaves  as a Maxwell-like tensor of rank $s - 2k$. In this sense the cubic vertices that we constructed in this paper automatically provide the general structure of possible interaction vertices also involving the traces of $\vf$, provided that one rescales the corresponding coefficients in $\cL_1$ from $s_i$ to $s_i - 2k_i$. Differences arise when computing the deformation of the gauge transformation and of the transversality constraint, however, due to the fact that the free Lagrangian is the one of $\vf$ while the equations of motion are those satisfied by  $\vf^{\, [k]}$.

In particular we know that the gauge variation of any cubic vertex,  say w.r.t. the first field $\vf_1$ of rank $s_1$, has to be of the form:
\be
\label{tracegeneric}
\d_0 \, \cL \, =\, -\, M*\, \d_1\, \vf_1 \, + \, \prd \prd\vf_1*\, \cC_1\,.
\ee
However, the traced equations of motion bring about a new contribution involving the double divergences of the fields:
\be
\label{traceM}
M^{\, [k]}\, =\, \Box\, \vf^{[k]}\, -\, \pr\, \prd\vf^{\, [k]}-2k\, \prd\prd\vf^{\, [k-1]}\, =\, 0\,,
\ee
so that, taking this contribution into account, one obtains
\be
\begin{split}
\label{tracespecific}
\d_0\, \cL \, =\, -\, M^{\, [k_1]}*\d_1\, \vf_1^{\, [k_1]}\, +\, \prd\prd\vf^{\, [k_1]}_1*\cC_1^{\, [k_1]}\, - \, 2k_1\,\prd\prd\vf^{\, [k_1 - 1]}_1*\d_1\, \vf^{[k_1]}\,,
\end{split}
\ee
where the functional forms of $\cC_1^{[k_1]}$ and $\d_1\vf^{[k_1]}_1$ are the same as in a standard vertex for tensors of rank $s_1-2k_1, \, s_2-2k_2$ and $s_3-2k_3$.

One has to keep in mind, however, that the traces are not actually independent fields and that one has  to bring \eqref{tracespecific} to the form \eqref{tracegeneric} inverting the trace operation. For instance
\be
- \, M^{[k_1]}*\d_1\, \vf_1^{\, [k_1]}\, \quad \longrightarrow \quad  -\, M^{(s_1)}*(\h^{k_1}\, \d_1\, \vf^{\, [k_1]})\, ,
\ee
with suitable combinatorial factors to be computed. Thus, in the vertices involving traces, the generalisation of the gauge transformation and of the deformation of the constraint read in general
\be
\begin{split}
&\d_1\, \vf^{(s_1)}\, = \, \r_{k_1, s_1} \, \h^{k_1}\, \d_1\, \vf^{[k_1]}_1\,,\\
&\prd \e^{(s_1-1)}\, =\, g\, \tfrac{1}{s_1(s_1ñ-1)}\, \left\{\tfrac{\r_{\, k_1, s_1 - 2}}{k_1+1}\, \h^{k_1}\, \cC_1^{[k_1]}\, -\, 2\, \r_{\, k_1-1, s_1-2} \,\h^{k_1-1} \, \d_1\vf^{[k_1]}_1\right\}\,.
\end{split}
\ee
where 
\be
\r_{k_1, s_1} \, := \, \frac{k_1!}{\prod_{i=1}^{k_1} \, \left({s - 2k_1 + 2i \atop 2}\right)} \, .
\ee
Including or not vertices of the schematic form
\be \label{Vk}
\cV \, \sim \, \pr^{\, n} \, \vf^{\, [k_1]}_1 \, \vf^{\, [k_2]}_2 \, \vf^{\, [k_3]}_3 \, ,
\ee
where $\vf^{\, [k_i]}_i$ denotes the $k-$th trace of a  tensor of rank $s_i$, would change the theory, of course. More generally, the cubic vertex involving particles whose polarisations are carried by the transverse-traceless parts of the traces in \eqref{Vk} may receive in our setting a number of independent contributions, namely all those contained in the vertices of the form
\be \label{Vki}
\cV \, \sim \, \pr^{\, n} \, \vf^{\, [k_1 - i_1]} \, \vf^{\, [k_2 -  i_2]} \, \vf^{\, [k_3 -  i_3]} \, ,
\ee
with $i_\ell = 0, 1, \ldots, k_\ell$. Whenever derivatives match one can simply consider the sum of all the contributions as the effective vertex for the three given particles, whose overall coefficient would be arbitrary, at this order. Let us also stress that, for independent vertices involving traces like \eqref{Vki}, the allowed number of derivatives turns out to be bounded as in \eqref{bound}:
\be \label{boundk}
\tilde{s}_1 \, + \, \tilde{s}_2 \, + \, \tilde{s}_3 \, - \, 2 \mbox{min}\{\tilde{s}_1,\, \tilde{s}_2,\, \tilde{s}_3 \} \, 
\leq \, \# \, \pr \, \leq \,  + \, \tilde{s}_1\, + \tilde{s}_2 \, + \, \tilde{s}_3 \, ,
\ee
with $\tilde{s}_{\ell} =  s_{\ell} - 2k_{\ell} + 2i_{\ell}, \, \ell = 1, 2, 3.$ 

The latter observation is relevant in order to assess the physical meaning of these additional interactions. In fact, the effective vertices for a given set of particles, subsumed in higher-rank tensor couplings, may  be above the derivative bound (6.12) for lower helicity modes. For instance, in a self-interacting vertex involving a rank-eight tensor $\vf_{\, \m_1 \, \ldots \, \m_8}$ saturating the lower bound in \eqref{bound} there will be eight derivatives; as a consequence,  the effective self-interacting spin-two vertex there included, via the couplings of the triple traces $\vf^{\, [3]}{}_{\, \m_1 \m_2}$, will also contain eight derivatives, which is above the {\it upper} bound for a triple of spin-two particles. For this reason, in a minimal theory where no traces are included, the self-interaction of the spin$-$two modes in $\vf_{\, \m_1 \, \ldots \, \m_8}$ would amount to a field redefinition, at cubic order. On the other hand, the inclusion of couplings among traces, like \eqref{Vki}, allows for the possibility of cubic interactions respecting \eqref{bound} for all the modes present in the spectrum.

The quartic step in the Noether procedure is usually assumed to constrain the relative coefficients of the cubic vertex. Results in this respect were recently discussed in \cite{Sleight:2016dba} from a holographic perspective and compared to a previous light-cone analysis performed in \cite{Metsaev:1991mt, Metsaev:1991nb}. Those results, however, refer to higher-spin theories whose spectra involve only one particle for each value of the spin (or possibly a finite number of them, taking into account that the corresponding results \cite{Metsaev:1991nb,Gwak:2015vfb,Gwak:2015jdo} also hold in the presence of Chan-Paton factors), which is generically not the case for Maxwell-like theories. Thus, while one may still expect that for any given theory the analysis of quartic vertices would allow to fix all the relative coefficients in the cubic couplings, it is not straightforward that the results obtained so far should apply to class of theories discussed in this work, in force of their different particle content.

\newpage

\section{(Anti) de Sitter cubic vertices} \label{sec: AdS}

Differently from Minkowski space time, the intrinsic infrared cut-off provided by a cosmological constant renders (Anti-) de Sitter backgrounds  a better framework where to explore the consistency of higher-spin  interactions. With respect to the case of flat background, the non-commutativity of the derivatives introduces an additional layer to the complexity of the problem and indeed, to our knowledge, metric-like (A)dS cubic interactions for arbitrary spins are known only at the TT  level or anyway in some partially gauge-fixed version \cite{Joung:2011ww, Manvelyan:2012ww, Sleight:2016dba}. In our opinion, the problem of uncovering their full off-shell structure is a relevant question to be addressed.\footnote{In the frame-like approach off-shell cubic vertices on AdS backgrounds were constructed in four dimensions by Fradkin and Vasiliev in \cite{Fradkin:1986qy}. See also \cite{Vasilev:2011xf, Boulanger:2012dx} for recent related constructions.}

The simplicity of the flat-space cubic vertex that we have obtained in Section \ref{sec: cubic} suggests that the Maxwell-like formulation may render the corresponding task more accessible also in presence of a cosmological constant. The Maxwell-like free Lagrangian in (A)dS was found in \cite{ML}, both for the irreducible and the reducible cases, and looks
\be \label{lag_ads_symm}
\cL \,=\, \12 \ \vf \,M_L\, \vf \, ,
\ee
where 
\be \label{M_L_symm}
M_L\, \vf \,\equiv\, M\, \vf \, - \, \fr{1}{L^{\, 2}} \,\left\{\, \left[\,(s - 2)(D + s - 3) \, - \, s\,\right] \, \vf \, - \, 2 \, g \, \vf^{\, \pe} \,\right\} \, ,
\ee
with the trace term absent in the irreducible case. In the present context, however, we found it useful to work in the ambient space formalism, so as to better exploit some of the useful properties of the Maxwell-like formulation in flat space, such as the absence of traces in  the free equations of motion. By means of these techniques we are able to construct fully off-shell cubic interactions for any triple of tensors of ranks $s_1, s_2$ and $s_3$. The price to pay is that the corresponding vertex structure gets more implicit.
 
 Our results apply to both the fully reducible and the fully irreducible cases directly, and we expect the same construction to work for the partially reducible cases as well.

In this section we follow the notation and the conventions of \cite{Joung:2011ww}.

\subsection{Ambient space formulation}

We introduce a master field in the ambient space
\be
\F(U,X)=\sum_{s=0}^{\infty} \frac{1}{s!}\F^{(s)}_{M_1\dots M_s}\, (X) \, U^{M_1}\dots U^{M_s}\, (X),
\ee
subject to homogeneity and tangentiality conditions:
\be
 (X\cdot\pr_X-U\cdot\pr_U+2)\F(U,X)=0\, , \quad (X\cdot\pr_U)\F(U,X)=0\,.
\ee
An appropriate modification of the flat-space measure is understood. To our purposes we only need to take into account that total derivatives are no longer negligible, namely
\be
\int d^D\m\,\,\pr_x\cdot(\dots)=\int d^D\m\,\l\, x\cdot(\dots)\,,
\ee
where $d^D\m$ is the new measure with while $\l$ is a constant directly related to the cosmological constant $\L$ of (A)dS$_d$.  In this setting, the free Maxwell-like action can be written in terms of derivative operators,
\be
\lag=\int \d^D\m\,e^{\pr_U\pr_V}[\Box_1-(U\pr_1)(\pr_U\pr_1)]\F(U,X_1)\F(V,X_2)\,,
\ee
where we have defined the new measure $\d^D\m=d^D\m\,d^DU\,d^DV\,\d(X_1-X_2)\,\d(U)\,\d(V)$. Moreover, we  introduce a master gauge parameter,
\be
E(U,X)=\sum_{s=0}^{\infty} \frac{1}{(s-1)!}E^{(s-1)}_{m_1\dots m_{s-1}}U^{m_1}\dots U^{m_{s-1}}\,,
\ee
encoding the full set of  transverse gauge transformations,
\be
\d\F(U,X)=(U\pr_X)\,E(U,X)\q,\q (\pr_U\pr_X)E(U,X)=0\, ,
\ee
while also satisfying appropriate homogeneity and tangentiality conditions
\be
(X\cdot\pr_X-U\cdot\pr_U)E(U,X)=0\q,\q (X\cdot\pr_U)E(U,X)=0\,.
\ee

\subsection{The AdS cubic vertex}

We want to write the most general cubic deformation of the free action that is consistent with the on-shell gauge invariance:
\be
\lag_1=\int \d^D\m\,C(\pr_i,\pr_{U_i})\F_1\F_2\F_3\,,\q \d_0 \lag_1\approx 0\,.
\ee
General arguments of symmetry and covariance lead us to state that the vertex function can be chosen to depend on the following six operators
\be
Y_i=\pr_{U_i}\pr_{i+1}\, , \q,\q Z_{i}=\pr_{U_{i-1}}\pr_{U_{i+1}}\,, \q i=1, 2, 3\, ,
\ee
where we have cyclicly identified $i=i+3$. Going off-shell requires the introduction of divergence and trace operators:
\be
(\pr_{U_i}\pr_{X_i})=\cD_i \,,\q(\pr_{U_i}\pr_{U_i})=\cT_i\,.
\ee
Thus, one looks for the most general operator $C(Y_i,Z_i)$ satisfying:
\be
\d_0\left\{\int \d^D\m\,C(Y_i,Z_i)\F_1\F_2\F_3\right\}\approx \cO (\cD_i,\cT_i)\,.
\ee
For instance, varying with respect to $\F_1$ we get
\be
\label{amb1}
\d_1\lag_{TT}=\int \d^D\m\,C(Y_i,Z_i)(U_1\pr_1)E_1\F_2\F_3=\int \d^D\m\,[C(Y_i,Z_i),(U_1\pr_1)]\,E_1\F_2\F_3\,.
\end{equation}
Following the same procedure outlined in Appendix \ref{B}, we can rearrange the commutator in \eqref{amb1} in such a way to recognise the part containing just TT operators. The vanishing  of this traceless-transverse sector imposes a differential constraint on the function $C$,
\be
\label{Cdef}
\left\{Y_2\pr_{Z_3}-Y_3\pr_{Z_2}-\l(Y_2\pr_{Y_2}-Y_3\pr_{Y_3})\pr_{Y_1} \right\}C(Y_i,Z_i)=0\, ,
\ee
whose general solution reads \cite{Joung:2011ww}:
\be
\begin{split}
&C(Y_i,Z_i)=e^{\l F(Z_i)}K(Y_i,G)|_{G\equiv G(Y,Z)}\,,\\
&G(Y_i,Z_i)=Y_1Z_1+Y_2Z_2+Y_3Z_3\,,\\
&F(Z_i)=Z_1\pr_{Y_2}\pr_{Y_3}+Z_1Z_2\pr_{Y_3}\pr_G+\text{cyclic perm.}+Z_1Z_2Z_3\pr_G^2\,.
\label{TTVertex}
\end{split}
\ee
In particular, the function $K(Y_i,G)$ is arbitrary and admits the Taylor expansion
\be
K(Y_i,G)=\sum_{s_1,s_2,s_3,n} g_{s_1,s_2,s_3,n}Y_1^{s_1-n}Y_2^{s_2-n}Y^{s_3-n}_3G^n\,,
\ee
whose coefficients $g_{s_1,s_2,s_3,n}$ that are nothing but the coupling constants of the vertices. Thus,  to any choice of  $K$ it corresponds a choice of the coupling constants\footnote{As a consequence, one can expect that the form of the function $K$ be fixed at the quartic order in the Noether procedure.}, while the parameter $n$ counts the number of derivatives in each vertex as follows
\be
\#\pr=s_1+s_2+s_3-2n\,. 
\ee
We can observe that, since $Y's$ and $G$ are differential operators, perturbative locality requires that
\be
n-\text{min}\{s_i\}\leq0\leq n \q \rightarrow \q s_1+s_2+s_3-2\text{min}\{s_i\}\leq\#\pr\leq s_1+s_2+s_3\,,
\ee
thus implying the usual flat-space bound. Finally, we observe that the function $e^{\l F(Z)}$ produces low-derivative terms and takes into account part of the low-derivative dimension-dependent tail typical of the AdS vertices.

The variation of the transverse and traceless part of the Lagrangian produces some terms containing next-order operators 
\be
\begin{split}
\d_0\lag_{TT}=\int\d^D\m\,\left\{ \tfrac{1}{2}(\Box_3-\Box_2-\Box_1)\pr_{Y_1}C-\cD_2\pr_{Z_3}C \right\}E_1\F_2\F_3\,.
\end{split}
\ee
As usual, we can now rearrange the $\Box$-terms in such a way to extract a part proportional to the equations of motion together with a part depending on the divergence of the fields:
\be
\begin{split}
\d_0\lag_{TT}=\int \d^D\m&\left[+\pr_{Y_1}C\mezzi(\cM_3-\cM_2)+\mezzi\pr_{Y_1}C\left(a_3\pr_3\cD_3-a_2\pr_2\cD_2\right)+\right.\\
&\left.-\cD_2\pr_{Z_3}C-\mezzi\pr_{Y_1}C\Box_1\right]E_1\F_2\F_3\,.
\end{split}
\ee
The last term suggests the form of the $\cD$-counterterm to be added to the TT Lagrangian
\be
\lag_{\cD}=\mezzi\int\d^D\m\,\left[\cD_1\pr_{Y_1}+\cD_2\pr_{Y_2}+\cD_{3}\pr_{Y_3}\right]C(Y_i,Z_i)\F(a_1,x_1)\F(a_2,x_2)\F(a_3,x_3)\,.
\ee
Proceeding along the same chain of compensations implemented in Section \ref{sec: cubic_red} one can reach the full off-shell completion of the vertex
\be
\label{lag-off}
\lag_1=\tfrac{1}{2}\int\d^D\m\left(2+\sum_i\cD_i\pr_{Y_i}+\sum_{i<j}\cD_i\cD_j\pr_{Y_i}\pr_{Y_j}+\cD_1\cD_2\cD_3\pr_{Y_1}\pr_{Y_2}\pr_{Y_3} \right)C(Y_i,Z_i)\F_1\F_2\F_3\,.
\ee
%

\subsection{Deformation of the constraint}

As we observed in Section \ref{sec: gauge4}, the Maxwell-like Noether procedure entails a subtlety, {\it i.e.} the variation of the Lagrangian \eqref{lag-off} vanishes only up to some weird contributions of the form
\be
\d_0\lag_1=\int\d^D\m\,\tfrac{1}{2}\left\{ -\pr_{Z_3}\pr_{Y_2}C-\cD_3\pr_{Y_3}\pr_{Z_3}\pr_{Y_2}C \right\}\left\{(\cD_2)^2+\l\cT_2\right\}E_1\F_2\F_3\, ,
\ee
where the presence of the double divergences points to the same mechanism of deformation of the constraint that we have already discussed in the flat case. In fact we can observe that, relaxing transversality constraint on the parameter, the variation of the free Lagrangian reads: 
\be
\d_0\lag_0=-2\int\d^D\m\,e^{\pr_{a_1}\pr_{a_2}}\left\{\cD_2^2+\l \cT_2\right\}\cD_1 E_1\F_2\,,
\ee
from which it is apparent that we can exactly compensate the $(\cD_2^2+\l\cT_2)$-terms appearing in $\d_0\lag_1$ by means of an appropriate deformation of the transversality constraint.

\subsection{Reduction to AdS covariant expressions: the $4-0-0$ example}

From the ambient-space vertex it is possible to recover an explicit AdS-covariant expression using the appropriate dictionary. The master field must be reduced to its AdS form,
\be
\vf=\sum_{s=0}^{s=\infty}\dfrac{1}{s!}\vf_{\m_1\dots\m_s} u^{a_1} e^{\m_1}_{a_1}\dots u^{a_s}e^{\m_s}_{a_s}\,,
\ee
where $u^{\m}$ lives on the tangent space and $e^\m_a$ is the AdS vielbein. Moreover the role of the partial derivatives $\pr_U$ is taken by the generalised covariant derivatives
\be
D_{\m}\,:=\, \nabla_\m+\dfrac{1}{2}\o^{ab}_\m u_{[a}\pr_{u^b]}
\ee
where $\nabla_\m$ is the usual covariant derivative on AdS. The coordinates $X^M$ of the embedding in $\mathbb{R}^{D}$ can be rearranged in such a way to extract a radial component
\be
X=R \hat X, \q \sqrt{X^2}=R
\ee
so that in particular the metric assumes the following form
\be
ds^2_{amb}=\h_{MN}dX^M dX^N=dR^2+\dfrac{R^2}{L^2}g_{\m\n}(x)dx^{\m}dx^{\n}=\tilde e^R \tilde e^R+\h_{ab}\tilde e^a\tilde e^b\,,
\ee
where $x$ now labels AdS coordinates while $g$ is the AdS metric, related to the embedding coordinates by
\be
g_{\m\n}= \dfrac{\pr X^M(x)}{\pr x^\m}\dfrac{\pr X_M(x)}{\pr x^\n}\,.
\ee
In this frame, the components of the ambient vielbein read
\be
\tilde e^R_R(R,x)=1, \q \tilde e^R_\m(R,x)=0,\q \tilde e^a_R(R,x)=0,\q \tilde e^a_\m(R,x)=\dfrac{R}{L}e^a_\m(x)\,.
\ee
In the same fashion, the tangent vectors split in a radial component and $d$ AdS components:
\be
U^M=\tilde e^M_R(R,x)v+\tilde e^M_a(R,x)u^a=\hat X^M(x)v+L\dfrac{\pr\hat X^M(x)}{\pr x^\m} e^\m_a(x)u^a\,.
\ee
Given this decomposition, we rearrange the ambient vertex operators as follows:
\be
\begin{split}
\label{reduced-op}
&\pr_{X^M}=\hat X^M\pr_R+\dfrac{L^2}{R}\dfrac{\pr\hat X_M}{\pr x^\m}[D_\m+\dfrac{1}{L}(u\cdot e_\m\pr_v-v\pr_u\cdot e_\m)]\,,\\
&\pr_{U^M}=\hat X_M\pr_v+L\dfrac{\hat X^M}{\pr x^\m}\pr u\cdot e_\m\,,
\end{split}
\ee
while the measure has to be understood as
\be
\int \dfrac{d^{d+1}X}{L} \d\left(\frac{\sqrt{X^2}}{L}-1\right)\, ,
\ee
where the delta function enforces the fields to have support  on the AdS hyper-surface in $\mathbb{R}^D$. 
Finally, we mention that the master field depends homogeneously on $R$
\be
\F(R,x;v,u)=\left(\dfrac{R}{L}\right)^{\D_h}\vf(x,u)
\ee
where $\D_h=u\cdot\pr_u-2$ is the homogeneity factor. 

As a concrete illustration we specialise \eqref{lag-off} to the case of a  $4-0-0$ vertex:
\be
\lag_{{\text 4-0-0}}= g_{\text 4-0-0} \left(Y_1^4+2\cD_1 Y_1^3\right)\cQ(X_1,U_1)\f_2(X_2,U_2)\f_3(X_3,U_3)\, ,
\ee
where $\cQ$ denotes the spin-four field and $\f_i$ are the scalar fields. Making use of the relations \eqref{reduced-op} together with some suitable tricks \cite{}, after some algebra one finds that the vertex reads
\be
\begin{split}
S_{\text 4-0-0}=\,g_{\text 4-0-0}\int d^{d}x \sqrt{-g}[&+\cQ_{\m\n\r\s}\nabla^\m \nabla^\n \nabla^\r \nabla^\s\f_2\f_3+2\nabla\cdot\cQ_{\m\n\r}\nabla^\m \nabla^\n \nabla^\r\f_2\f_3+\\
&+\L^2(4\cQ'_{\m\n}\nabla^\m \nabla^\n\f_2\f_3-6\nabla\cdot\cQ'_\r \nabla^\r\f_2\f_3)+\\
&-24\L^4\cQ''\f_2\f_3] \, ,
\end{split}
\ee
where in particular we can appreciate the appearance of a contribution in the double trace of the spin-four field, that would not be present in the irreducible (nor in the Fronsdal case), and thus represents a peculiarity of the reducible formulation.

\subsection{Off-shell generating functions and Grassmann variables}

We would like to notice here some interesting aspects of the off-shell cubic vertices, {\it i.e.} vertices involving fields that are not subject to any conditions like transversality and tracelessness, nor involve partially gauge fixed fields. For the Fronsdal fields, the off-shell generating functions were found to have a simple form \citep{Manvelyan:2010je,Mkrtchyan:2011uh}. It was noticed in \citep{Manvelyan:2010je}, that certain deformations of the TT vertex involving Grassmann variables allow to immediately write down the generating functions for off-shell vertices. The origin of (BRST-like) Grassmann structure appearing in these generating functions is not well understood.

It turns out that the Maxwell-like vertices follow the same pattern and  thus their off-shell formulation can be achieved in a similar way. Remarkably, the procedure for  Maxwell-like fields requires a deformation only  involving the derivative operators $Y_i$ but not the contractions $Z_i$. This is related to the fact that traces are not appearing in the Maxwell-like vertices.  Indeed in \citep{Manvelyan:2010je,Mkrtchyan:2011uh} the contractions comprise Grassmann deformations involving traces of the fields. One can check easily, that the off-shell vertices \eqref{lag-off} can be written in the following form
\be
\label{gf-off}
\lag_1=\int\d^D\hat\m\, C(\hat Y_i,Z_i)\F_1\F_2\F_3\, ,
\ee
 where $C(Y,Z)$ is the function \eqref{TTVertex}, that defines the TT vertices, while
\bea
&\hat Y_i &= \, Y_i+\frac12\eta_i\bar\eta_{i+1}\cD_{i+1}-\frac12\eta_i\bar\eta_{i-1}\cD_{i-1}
				+\frac12\eta_i\bar\eta_{i}\cD_{i}\,,\\
&\d^D\hat\m& = \, \d^D\m \Pi_{i=1}^3 d\eta_i d\bar\eta_i\, e^{\sum_{i=1}^{3}\eta_i\bar\eta_i}\,.
\eea
provide suitable deformations of the operators $Y_i$ and of the integration measure, involving  three pairs of Grassmannian variables $\eta_i\,,\,\bar\eta_i$\,, $i=1,2,3$\,, one pair for each field.
As shown in \citep{Manvelyan:2010je} a similar  procedure works for both the anti-symmetric ansatz and the cyclic one. Since there is no conceptual difference between different ans\"atze, we expect it to work for Maxwell-like fields on flat and AdS backgrounds too.
In fact, based on this evidence one can also conjecture, that the Fronsdal off-shell vertices in AdS can be  obtained through the same procedure as in \citep{Manvelyan:2010je} starting from the TT vertices of \citep{Joung:2011ww}, and can be simply written as:
\be
\label{gf-off:Fronsdal}
\lag^{\, \cF}_1=\int\d^D\hat\m\, C(\hat Y_i, \hat Z_i)\F_1\F_2\F_3\,,
\ee
with the only modifications compared to  \eqref{gf-off} being that $\cD_i$ is to be interpreted as the full de Donder tensor, while 
\be
\hat Z_i = Z_i+\frac12 \eta_{i+1}\bar \eta_{i-1} {\cal T}_{i-1}\,.
\ee
In particular, in the de Donder gauge we recover the gauge-fixed result of \cite{Sleight:2016dba}.

\section{Outlook} \label{sec: out}

In this work we constructed cubic interactions for massless fields of any spin described at the free level by the Maxwell-like Lagrangians \eqref{ML}. This formulation is more flexible than the Fronsdal one, in that it can accommodate all possible unitary spectra of particles whose degrees of freedom are carried by symmetric tensors: from the maximally reducible representation associated to traceful fields and parameters to the single-particle description for the fully traceless case, all intermediate cases appear to be admissible and can be conveniently described in our approach, with relatively minor modifications in passing from one case to the other. Our construction covers indeed all possible spectra for the case of flat-space cubic vertices, while also extending to the case of (A)dS backgrounds.

One peculiarity of the Maxwell-like theories, as we stressed, is the spectrum that they would describe in a putative complete non-linear construction. As we saw, from the perspective of the solution to the Noether conditions reducibility of the spectra does not appear to play any significant role. In this sense, one would expect that the quartic-level analysis reproduce the main features of the Fronsdal case, including in particular the need for infinitely-many tensors of unbounded ranks. Assuming that a complete theory exist, possibly on (A)dS, involving for simplicity one copy for each even-rank Maxwell-like tensor, the resulting particle spectrum in the fully reducible case would be of the following type (where on each column one can read the particle content associated to  the tensor of the corresponding rank):
\begin{table}[h]
\centering
\begin{tabular}{|c|ccccc|}
\hline
\diagbox{spin}{rank} &\ 0 \ & \ 2 \ & \ 4 \ & \ 6 \ & \ \dots \\
\hline
 0 & {\color{blue} x} & {\color{blue} x} & {\color{blue} x} & {\color{blue} x} & {\color{blue} \dots} \\ 
 2 &  & {\color{blue} x} & {\color{blue} x} & {\color{blue} x} & {\color{blue} \dots} \\ 
 4 &  &  & {\color{blue} x} & {\color{blue} x} & {\color{blue} \dots} \\ 
 6 &  &  &  & {\color{blue} x} & {\color{blue} \dots} \\ 
\vdots &  &  &  & & ${\color{blue} \ddots}$ \\ 
\hline
\end{tabular}
\end{table}
\noindent with infinitely-many particles present for any given spin, although with seemingly decreasing multiplicity as the spin increases. Let us stress that there would be only one ``graviton'' in the spectrum, {\it i.e.} only one minimally coupled spin$-2$ particle, all the other spin$-2$ particles present in the spectrum only entering non-minimal couplings. 

Other particle contents can be realised upon enforcing trace constraints of various sorts. Many of them would give rise anyway to spectra involving infinitely-many particles with given spins, whenever the reducibility conditions be chosen so as to allow for a number of particles that increases with the rank of each tensor. For instance one may think of truncating away all the scalars (probably with the exception of the one carried by the rank$-0$ tensor) or all the spin$-2$ particles (probably with the exception of the one carried by the rank$-2$ tensor), and so forth.  

Moreover, in a theory with infinitely-many Maxwell-like fields the very notion of ``multiplicity''  of a given spin would be no longer obviously defined, and the picture sketched above may turn out to be even misleading, to some extent. For instance, Let us consider one copy of each spin described by reducible Maxwell-like fields. As already noticed, at first sight it may seem that the would-be theory of fully reducible Maxwell-like higher spins would contain in its spectrum more spin$-s_1$ fields than spin$-s_2$ fields for $s_1 < s_2$. This might not necessarily be the case, though. In Figure \ref{fig1} below\footnote{We use the notation $Vas_d$ for the Vasiliev algebra in $(A)dS_d$ which, is also known as Eastwood-Vasiliev algebra and denoted in literature as $hso[d-1,2]$ for negative cosmological constant. We do not need to specify the sign of the cosmological constant, therefore here $Vas_d$ is some real form of the complex algebra $hs(so_{d+1})$, studied in \citep{Joung:2014qya}\,.}, we show how could a theory with equal number of particles of any spin be organized into Maxwell-like multiplets.
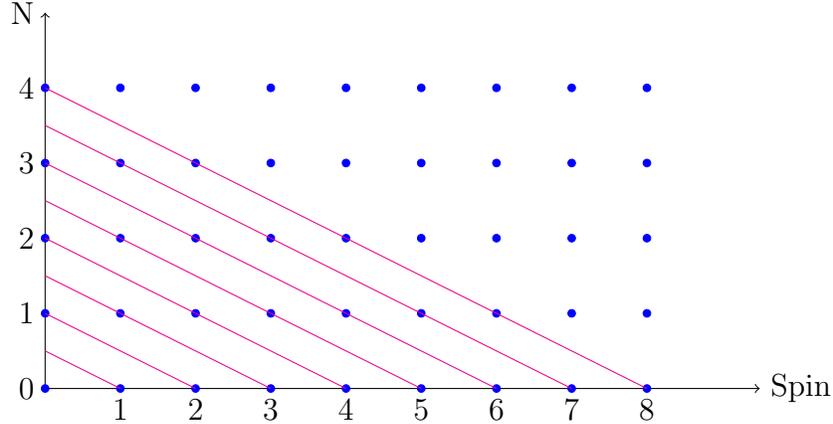
\begin{figure}[h]
\centering
\begin{tikzpicture}
\draw [->] (0,0) -- (9.5,0);
\node [right] at (9.5,0) {Spin};
\draw [->] (0,0) -- (0,5);
\node [left] at (0,5) {N};
\draw [blue,fill=blue] (0,0) circle(.05);
\draw [blue,fill=blue] (0,1) circle(.05);
\draw [blue,fill=blue] (0,2) circle(.05);
\draw [blue,fill=blue] (0,3) circle(.05);
\draw [blue,fill=blue] (0,4) circle(.05);
\node [left] at (0,0) {$0$};
\node [left] at (0,1) {$1$};
\node [left] at (0,2) {$2$};
\node [left] at (0,3) {$3$};
\node [left] at (0,4) {$4$};
\draw [blue,fill=blue] (1,0) circle(.05);
\draw [blue,fill=blue] (2,0) circle(.05);
\draw [blue,fill=blue] (3,0) circle(.05);
\draw [blue,fill=blue] (4,0) circle(.05);
\draw [blue,fill=blue] (5,0) circle(.05);
\draw [blue,fill=blue] (6,0) circle(.05);
\draw [blue,fill=blue] (7,0) circle(.05);
\draw [blue,fill=blue] (8,0) circle(.05);
\node [below] at (1,0) {$1$};
\node [below] at (2,0) {$2$};
\node [below] at (3,0) {$3$};
\node [below] at (4,0) {$4$};
\node [below] at (5,0) {$5$};
\node [below] at (6,0) {$6$};
\node [below] at (7,0) {$7$};
\node [below] at (8,0) {$8$};
\draw [blue,fill=blue] (1,1) circle(.05);
\draw [blue,fill=blue] (2,1) circle(.05);
\draw [blue,fill=blue] (3,1) circle(.05);
\draw [blue,fill=blue] (4,1) circle(.05);
\draw [blue,fill=blue] (5,1) circle(.05);
\draw [blue,fill=blue] (6,1) circle(.05);
\draw [blue,fill=blue] (7,1) circle(.05);
\draw [blue,fill=blue] (8,1) circle(.05);
\draw [blue,fill=blue] (1,2) circle(.05);
\draw [blue,fill=blue] (2,2) circle(.05);
\draw [blue,fill=blue] (3,2) circle(.05);
\draw [blue,fill=blue] (4,2) circle(.05);
\draw [blue,fill=blue] (5,2) circle(.05);
\draw [blue,fill=blue] (6,2) circle(.05);
\draw [blue,fill=blue] (7,2) circle(.05);
\draw [blue,fill=blue] (8,2) circle(.05);
\draw [blue,fill=blue] (1,3) circle(.05);
\draw [blue,fill=blue] (2,3) circle(.05);
\draw [blue,fill=blue] (3,3) circle(.05);
\draw [blue,fill=blue] (4,3) circle(.05);
\draw [blue,fill=blue] (5,3) circle(.05);
\draw [blue,fill=blue] (6,3) circle(.05);
\draw [blue,fill=blue] (7,3) circle(.05);
\draw [blue,fill=blue] (8,3) circle(.05);
\draw [blue,fill=blue] (1,4) circle(.05);
\draw [blue,fill=blue] (2,4) circle(.05);
\draw [blue,fill=blue] (3,4) circle(.05);
\draw [blue,fill=blue] (4,4) circle(.05);
\draw [blue,fill=blue] (5,4) circle(.05);
\draw [blue,fill=blue] (6,4) circle(.05);
\draw [blue,fill=blue] (7,4) circle(.05);
\draw [blue,fill=blue] (8,4) circle(.05);
\draw [magenta, thin] (1,0) -- (0,.5);
\draw [magenta, thin] (2,0) -- (0,1);
\draw [magenta, thin] (3,0) -- (0,1.5);
\draw [magenta, thin] (4,0) -- (0,2);
\draw [magenta, thin] (5,0) -- (0,2.5);
\draw [magenta, thin] (6,0) -- (0,3);
\draw [magenta, thin] (7,0) -- (0,3.5);
\draw [magenta, thin] (8,0) -- (0,4);
\end{tikzpicture}
\caption{Maxwell-like decomposition of the stringy massless spectrum of generalised Vasiliev theory with symmetry algebra $U(\infty)\otimes Vas_d$\,. $N$ numerates Regge trajectories, starting with the trajectory $N=0$. Blue dots with coordinates $(s,n)$ correspond to massless states of spin $s$ in $n$-th ``Regge trajectory'', while the lines connect the particles that combine into reducible representation described by Maxwell-like field.}
\label{fig1}
\end{figure}
For the case of a countable infinity of ``Regge trajectories'', which contain one massless particle of each integer spin, we arrive at a spectrum that could be described by one copy of fully reducible field of each spin. Now, imagine that the number of Regge trajectories is finite, say $M$. Then, we can still use the same Maxwell-like fields, all of which are subject to $M$-th traceless constraint, as shown in the Figure \ref{fig2}.  Complete, fully non-linear higher-spin theories with such kind of spectra have not been constructed so far, to the best of our knowledge, nor is it obvious that they should exist at all. The study of the global symmetries of the corresponding  models, a well as  the investigation about the existence of prospective holographic counterparts, provide conceivable complementary guiding lines along which to address the question about their existence.

In particular, in its conventional formulation, Vasiliev theory and its coloured generalisations involve finitely many fields of any given spin and thus correspond to the case of a finite number of Regge trajectories, whose spectra could be reproduced in terms of irreducible or partially reducible Maxwell-like fields.  Let us also notice that the Maxwell-like description implements a grading between even and odd spins, therefore the spectra of all known coloured Vasiliev theories, based on the algebras discussed in \citep{Konshtein:1988yg,Konstein:1989ij}, can still  be matched by Maxwell-like spectra.

\begin{figure}[h]
\centering
\begin{tikzpicture}
\draw [->] (0,0) -- (10.5,0);
\node [right] at (10.5,0) {Spin};
\draw [->] (0,0) -- (0,4);
\node [left] at (0,4) {N};
\draw [blue,fill=blue] (0,0) circle(.05);
\draw [blue,fill=blue] (0,1) circle(.05);
\draw [blue,fill=blue] (0,2) circle(.05);
\draw [blue,fill=blue] (0,3) circle(.05);
\node [left] at (0,0) {$0$};
\node [left] at (0,1) {$1$};
\node [left] at (0,2) {$2$};
\node [left] at (0,3) {$3$};
\draw [blue,fill=blue] (1,0) circle(.05);
\draw [blue,fill=blue] (2,0) circle(.05);
\draw [blue,fill=blue] (3,0) circle(.05);
\draw [blue,fill=blue] (4,0) circle(.05);
\draw [blue,fill=blue] (5,0) circle(.05);
\draw [blue,fill=blue] (6,0) circle(.05);
\draw [blue,fill=blue] (7,0) circle(.05);
\draw [blue,fill=blue] (8,0) circle(.05);
\draw [blue,fill=blue] (9,0) circle(.05);
\draw [blue,fill=blue] (10,0) circle(.05);
\node [below] at (1,0) {$1$};
\node [below] at (2,0) {$2$};
\node [below] at (3,0) {$3$};
\node [below] at (4,0) {$4$};
\node [below] at (5,0) {$5$};
\node [below] at (6,0) {$6$};
\node [below] at (7,0) {$7$};
\node [below] at (8,0) {$8$};
\node [below] at (9,0) {$9$};
\node [below] at (10,0) {$10$};
\draw [blue,fill=blue] (1,1) circle(.05);
\draw [blue,fill=blue] (2,1) circle(.05);
\draw [blue,fill=blue] (3,1) circle(.05);
\draw [blue,fill=blue] (4,1) circle(.05);
\draw [blue,fill=blue] (5,1) circle(.05);
\draw [blue,fill=blue] (6,1) circle(.05);
\draw [blue,fill=blue] (7,1) circle(.05);
\draw [blue,fill=blue] (8,1) circle(.05);
\draw [blue,fill=blue] (9,1) circle(.05);
\draw [blue,fill=blue] (10,1) circle(.05);
\draw [blue,fill=blue] (1,2) circle(.05);
\draw [blue,fill=blue] (2,2) circle(.05);
\draw [blue,fill=blue] (3,2) circle(.05);
\draw [blue,fill=blue] (4,2) circle(.05);
\draw [blue,fill=blue] (5,2) circle(.05);
\draw [blue,fill=blue] (6,2) circle(.05);
\draw [blue,fill=blue] (7,2) circle(.05);
\draw [blue,fill=blue] (8,2) circle(.05);
\draw [blue,fill=blue] (9,2) circle(.05);
\draw [blue,fill=blue] (10,2) circle(.05);
\draw [blue,fill=blue] (1,3) circle(.05);
\draw [blue,fill=blue] (2,3) circle(.05);
\draw [blue,fill=blue] (3,3) circle(.05);
\draw [blue,fill=blue] (4,3) circle(.05);
\draw [blue,fill=blue] (5,3) circle(.05);
\draw [blue,fill=blue] (6,3) circle(.05);
\draw [blue,fill=blue] (7,3) circle(.05);
\draw [blue,fill=blue] (8,3) circle(.05);
\draw [blue,fill=blue] (9,3) circle(.05);
\draw [blue,fill=blue] (10,3) circle(.05);
\draw [magenta, thin] (1,0) -- (0,.5);
\draw [magenta, thin] (2,0) -- (0,1);
\draw [magenta, thin] (3,0) -- (0,1.5);
\draw [magenta, thin] (4,0) -- (0,2);
\draw [magenta, thin] (5,0) -- (0,2.5);
\draw [magenta, thin] (6,0) -- (0,3);
\draw [magenta, thin] (7,0) -- (1,3);
\draw [magenta, thin] (8,0) -- (2,3);
\draw [magenta, thin] (9,0) -- (3,3);
\draw [magenta, thin] (10,0) -- (4,3);
\end{tikzpicture}
\caption{Maxwell-like decomposition of the spectrum of generalized Vasiliev theory with four Regge trajectories and symmetry algebra $U(2)\otimes Vas_d$\,. Blue dots with coordinates $(s,n)$ correspond to massless states of spin $s$ in $n$-th ``Regge trajectory'', while the lines connect the particles that combine into reducible representation described by Maxwell-like field with fourth trace constraint.}
\label{fig2}
\end{figure}
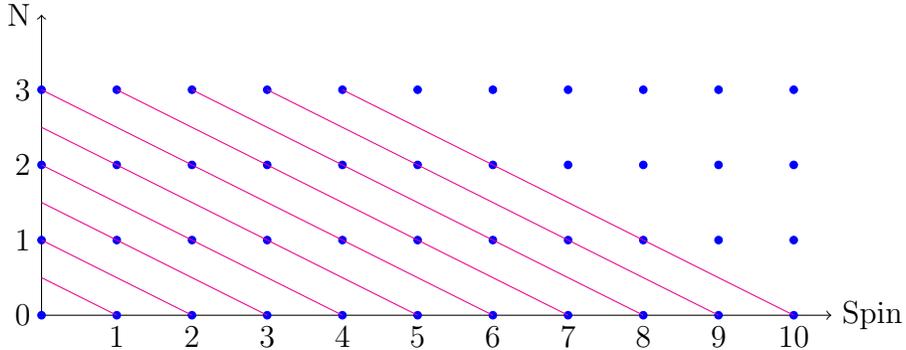
 It is worth recalling, however,  that the fully reducible Maxwell-like theories describe free  tensionless strings; from this perspective it is far from being apparent which possible truncations may or may not be allowed when interactions are turned on while the tension is still kept to zero. 
The Regge trajectories in String Theory contain massive states, and the collapse to zero mass of the full spectrum, including mixed-symmetry particles, is described in \cite{st}.  In the simplest case, the first Regge trajectory contains massive states corresponding to symmetric fields of arbitrary rank. In the massless limit each spin$-s$ massive particle decomposes into a tower of helicity states, from $0$ to $s$. One can actually think of this spectrum as composed of two towers of reducible massless fields, that can be Higgsed in the following way: a spin$-s$ Maxwell-like field from one tower eats a spin$-(s-1)$ Maxwell-like field from the other tower and becomes massive.

The fact that traces do not appear in the flat-space Maxwell-like action might turn out to carry a relevant piece of information about the full non-linear theory. It was recently argued  that the remarkable properties of gravitational amplitudes allowing to interpret them as squares of gauge theory amplitudes might be made manifest at the level of the action \citep{Cheung:2016say}. An important condition for this to be possible is that the action does not contain traces of the field. This is automatically satisfied for Maxwell-like fields of any spin without any gauge fixing. 

\acknowledgments

We are grateful to  A. Bengtsson, P. Benincasa, D. Blas, N. Boulanger, A. Campoleoni, E. Conde, E. Joung, R. Metsaev, S. J. Rey, A. Sagnotti, P. Sundell,  M. Taronna, M. Tsulaia and Yu. Zinoviev for useful discussions and comments. The work of D. F. was supported in part by Scuola Normale Superiore and INFN (I.S. Stefi). The work of K. M. was partly supported by the BK21 Plus Program funded by the Ministry of Education (MOE, Korea) and National Research Foundation of Korea (NRF) and by Alexander von Humboldt foundation. Both D.F. and K.M. were also supported in part by the Munich Institute for Astro- and Particle Physics (MIAPP) of the DFG cluster of excellence ``Origin and Structure of the Universe''.  K.M. would like to thank Scuola Normale Superiore and INFN Sezione di Pisa for the kind hospitality extended to him during the final stages of this work.

\appendix

\section{Notation and conventions}\label{A}

\subsection{Free fields}

For the case of free fields we exploit a notation where all  indices are omitted. Thus a rank-$s$ symmetric tensor $\vf_{\, \m_1 \, \ldots \, \m_s}$ is denoted simply as $\vf$, while $\vf^{\, \pe}$ denotes the Lorentz trace of $\vf$ and $\prd \vf$  stands for its divergence.  Products of different tensors are symmetrised with the minimal number of terms and with no weight factors. In particular the Abelian gauge transformation of free potentials is denoted simply as $\d \, \vf = \pr \, \e$, where $\e$ is the rank$-(s-1)$ gauge parameter. For more details about the combinatorics the reader can consult {\it e.g.} \cite{fs1}.

\subsection{Interacting fields}

In order to deal with cubic and higher-order vertices one needs a notation allowing to keep track of the various possible types of contractions \cite{Manvelyan:2010jr}. For a rank$-s$ tensor $\vf_{\, \m_1 \, \ldots \, \m_s}$ we exploit $s$ powers of an auxiliary commuting vector $a^{\m}$ of the tangent space at the base-point $x$ to define
\be
\vf^{(s)}\, (x; a)\, =\, \vf_{\m_{1}\cdots\m_{s}}\, (x) \, a^{\m_{1}}\cdots a^{\m_{s}}\, .
\ee
Derivations with respect to $x^{\, \m}$ or to  $a^{\, \m}$ are distinguished as follows:
\be
\fr{\pr}{\pr \, x^{\, \m}} \, := \, \pr_{\, \m}, \hskip 3cm \fr{\pr}{\pr \, a^{\, \m}} \, := \, \pr^{\, a}{}_{\m} \, .
\ee
Correspondingly, all the relevant operations to be performed on tensor fields, involving either space-time derivatives or  contraction and symmetrization of indices, are implemented in terms of differential operators. We collect the main ones in the following list, where we also display the proper conversion factors needed to compare with our index-free convention for free fields:
\begin{align}
&& && && && && &a \cdot \pr \, \vf^{(s)}\, (x; a) & & = & & \fr{1}{s + 1} \, (\pr \, \vf)^{\, (s + 1)} \, (x; a), & && && && &&\\
&& && && && && & (a \cdot \pr)^{\, 2} \,\vf^{(s)}\, (x; a) & & = & & \fr{1}{{s + 2\choose 2}}\, (\pr^{\, 2} \, \vf)^{\, (s + 2)} \,  (x; a), & && && && &&\\
&& && && && && &\prd \pr_a \,\vf^{(s)}\, (x; a)  & & = & &s \, (\prd \vf)^{\, (s-1)} \, (x; a), & && && && &&\\
&& && && && && & \Box_a \, \vf^{(s)}\, (x; a)  & & = & & s \, (s-1) \, (\vf^{\, \pe})^{\, (s-2)}\, (x; a), & && && && &&\\
&& && && && && & a^{\, 2} \, \vf^{(s)}\, (x; a)  & & = & & \fr{1}{{s + 2\choose 2}}\, (\h \, \vf)^{\, (s + 2)} \,  (x; a) \, .& && && && &&
\end{align}
In this notation, for instance, the Maxwell-like and Fronsdal tensors take the form
\be
\begin{split} \label{MF}
& M^{(s)}\, (x; a) \, = \, \left\{\Box \, - \, a \cdot \pr \, \prd \pr_a \right\}\,\vf^{(s)}\, (x; a)\, , \\
& \cF^{(s)}\, (x; a) \, = \, \left\{\Box \, - \, a \cdot \pr \, \prd \pr_a \,+ \, \12 \, (a \cdot \pr)^{\, 2} \, \Box_a \right\}\,\vf^{(s)}\, (x; a)\, . \\
\end{split}
\ee
Quadratic scalars can be computed by means of the contraction operator
\be
*_a \, := \, \fr{1}{(s \, !)^2} \, \prod_{i=1}^s \, \overleftarrow{\pr}^{a}{}_{\m_i} \, \overrightarrow{\pr}^{\m_i}{}_{a}\, ,
\ee
so that, for instance
\be
\begin{split}
& \vf \cdot \psi \, = \,  \vf^{(s)}\, (x; a) \, *_a \,  \psi^{(s)}\, (x; a)   ,\\
& \cL_{M} \, = \, \12 \, \vf \, M \, = \, \12\, \vf^{(s)}\, (x; a) \, *_a \, M^{(s)}\, (x; a)   ,\\
& \cL_{\cF} \, = \, \12 \, \vf \, \left\{\cF \, - \, \12 \, \h \, \cF^{\, \pe}\right\} \, = \,  \12\, \vf^{(s)}\, (x; a) \, *_a \, 
\left\{1 \, - \, \fr{1}{4} \,  a^{\, 2} \, \Box_a \right\} \cF^{(s)}\, (x; a)   . 
\end{split}
\ee
%

\section{Some useful commutators}\label{B}

In order to solve the Noether procedure at the cubic level we need an explicit expression for commutators of the following type
\be \label{comm}
[T\, (n_{1},n_{2},n_{3}|Q_{12},Q_{23},Q_{31})\, ,\, (a \cdot \pr_{1})]\, ,
\ee
where T is the operator defined in \eqref{T}, that we report here for convenience:
\be \label{Tapp}
T\,(Q, \, n)\, = \, (\pr_a \cdot \pr_2)^{\, n_1} \, (\pr_b \cdot \pr_3)^{\, n_2} \, (\pr_c \cdot \pr_1)^{\, n_3} \, (\pr_a \cdot \pr_b)^{\, Q_{12}} \, (\pr_b \cdot \pr_c)^{\, Q_{23}} \,  (\pr_c \cdot \pr_a)^{\, Q_{31}}\, ,
\ee  
from which one can read the operators that can give a contribution to \eqref{comm}. We shall use the general fact that, if two operators $\cA$ and  $\cB$ are such that $[[\cA,\cB],\cA]=0$, then one can demonstrate by induction the following relation
\be \label{AB}
[\cA^n,\, \cB]\, =\, n\, \cA^{n-1}\, [\cA,\, \cB]\,.
\ee
Using \eqref{AB} we can reduce the various non trivial commutators in \eqref{comm} to the following building blocks 
\be
\begin{split}
& [(\pr_a \cdot \pr_2)^{n_1},\, (a \cdot \pr_{1})]\, = \, \pr_1 \cdot \pr_2 \, ,    \\
& [(\pr_a \cdot \pr_b)^{Q_{12}}, \, (a \cdot \pr_{1})]   \, = \, - \, (\pr_b \cdot \pr_3)^{n_2}  \, - \, \pr_2 \cdot \pr_b  \, , \\
& [(\pr_c \cdot \pr_a)^{Q_{31}}, \, (a \cdot \pr_{1})]   \, = \, \pr_c \cdot \pr_1 \, ,
\end{split}
\ee
from which we can evaluate \eqref{comm}
\begin{align}
\nonumber
\left[T\, ,\, (a \cdot \pr_{1})  \right]=& +Q_{31}\, T(n_{3}+1\, | \, Q_{31} \, -\, 1) \, - \, Q_{12}T \, (n_{2}+1\, |Q_{12}-1) \\ 
& - \, Q_{12}\, T \, (Q_{12}-1) \, (\pr_{b} \cdot \pr_{2}) \, + \, n_{1}\, T\, (n_{1}-1)\, (\pr_{1} \cdot \pr_{2})\,,
\end{align}
and similarly for $[T\, ,\, (b \cdot \pr_{2})]$ and $[T\, ,\, (c \cdot \pr_{3})]$.

\section{On the need for deforming the constraint}\label{C}

One can argue that the setup in this work is limited in that we do not include more than one divergence of any given field in the cubic ansatz. Despite the fact that double (and higher) divergences are vanishing on the free mass shell, they are non-locally related to the equations of motion, and thus they cannot be removed by local field redefinitions. Therefore, the most general ansatz for the cubic vertex should involve higher divergences of the fields and allow for a richer parameter space for the cubic vertex, thus potentially altering our conclusion about the need to deform the constraints on the gauge parameters.

Now we will try to solve Noether procedure for a general ansatz involving higher divergences of the fields, 
\bea\label{L}
\cL_3&=& \sum_{m_1,m_2,m_3} \cL^{m_1,m_2,m_3}\,, \nonumber\\
\cL^{m_1,m_2,m_3} &=& \sum_{\a+\b+\g=n} C_{\a,\b,\g}^{m_1,m_2,m_3}T(\a,\b,\g|m_1,m_2,m_3)\vf^{s_1}(a_1)\vf^{s_2}(a_2)\vf^{s_3}(a_3)\,,
\eea
where the $m$'s can take values higher than one, and in general the variation of the (\ref{L}) should be proportional to the free equations of motion and to their divergences.

First we compute the divergences of the equations of motion for any field. The kinetic tensor in the language of the oscillators $a$ reads as in \eqref{MF}:
\be
M(\vf(a))=(\Box-(a\cdot\pr)(\pr\cdot\pr_{a}))\vf(a) \, .
\ee
Now we compute the n-th divergence of the equation of motion:
\be
(\pr\cdot\pr_a)^n M(\vf(a))=[(1-n)\Box(\pr\cdot\pr_a)^n-(a\cdot\pr)(\pr\cdot\pr_{a})^{n+1}]\vf(a)\, .
\ee

 The general solution should be gauge invariant with respect to the gauge variation of any of the fields. We are looking for a universal solution to the Noether equation, which should be symmetric with respect to the three fields included. We therefore check the gauge invariance with respect to the variation of only one of the fields, namely $\vf^{\, (s_1)}$. Taking into account that the terms with higher than one divergences on the field $\vf^{\, (s_1)}$ do not contribute to the gauge variation of cubic Lagrangian due to transversality constraint\footnote{If we allow deformation of the transversality constraint, the variation of the terms with more than one divergence on the field $\vf^{\, (s_1)}$ contribute to orders higher than the cubic one. In any case we can drop them while discussing the cubic order only.}, the solution of the Noether equation ensuring gauge invariance with respect to the gauge transformation of the field $\vf^{\,(s_1)}$ can give information only about the coefficients $C^{0,m_2,m_3}_{\a,\b,\g}, C^{1,m_2,m_3}_{\a,\b,\g}$. From the symmetry of the general solution with respect to three fields, we should be able to recover also other coefficients different from zero, if any. The variation of the Lagrangian (\ref{L}) gives (the commutators of the new $T$-operator can be computed as in Appendix \ref{B}):
\bea
\d_0 \cL^{0,m_2,m_3}=\sum_{\a+\b+\g=n}C^{0,m_2,m_3}_{\a,\b,\g}
\{(s_1-n+\a)T(\a,\b,\g|0,m_2,m_3)(\Box_3-\Box_2)\nonumber\\
-\frac{1}{2}\g T(\a,\b,\g-1)|0,m_2+1,m_3)-\frac{1}{2}\g T(\a,\b,\g-1)|0,m_2,m_3)\nonumber\\
-\frac{1}{2}\b T(\a,\b-1,\g|0,m_2,m_3+1)+\frac{1}{2}\b T(\a,\b-1,\g)|0,m_2,m_3) \}\, , \nonumber\\
\L^{s_1-1}(a)\vf^{s_2}(b)\vf^{s_3}(c)\, , \label{varL0}\\
\d_0 \cL^{1,m_2,m_3}=\sum_{\a+\b+\g=n}C^{1,m_2,m_3}_{\a,\b,\g}T(\a,\b,\g|0,m_2,m_3)
\Box_1\L^{s_1-1}(a_1)\vf^{s_2}(a_2)\vf^{s_3}(a_3)\, .\label{varL1}
\eea
Now let us  introduce two groups of compensating terms:
\bea
\D^{0,m_2,m_3}_2=\sum_{\a,\b,\g} B^{0,m_2,m_3}_{\a,\b,\g}\hat T(\a,\b,\g|0,m_2,m_3)(a_2\cdot\pr_2)
(\pr_{a_2}\cdot \pr_{2})^{m_2}(\pr_{a_3}\cdot \pr_{3})^{m_3} \, ,
\nonumber\\
\L^{s_1-1}(a)\vf^{s_2}(b)\vf^{s_3}(c),\label{defB}\\
\D^{0,m_2,m_3}_3=\sum_{\a,\b,\g} D^{0,m_2,m_3}_{\a,\b,\g}\hat T(\a,\b,\g|0,m_2,m_3)(a_3\cdot\pr_3)
(\pr_{a_2}\cdot \pr_{2})^{m_2}(\pr_{a_3}\cdot \pr_{3})^{m_3}\, ,
\nonumber\\
\L^{s_1-1}(a_1)\vf^{s_2}(a_2)\vf^{s_3}(a_3)\, ,\label{defD}
\eea
where\footnote{Using $\hat T$ we just normal order $(a_2\cdot\pr_2), (a_3\cdot\pr_3)$ with the divergences $(\pr_{a_2}\cdot\pr_2), (\pr_{a_3}\cdot\pr_3)$.}
\bea
\hat T(\a,\b,\g|m_1,m_2,m_3)=&(\pr_{a_1}\cdot\pr_{23})^{s_1-m_1-n+\a}(\pr_{a_2}\cdot \pr_{31})^{s_2-m_2-n+\b}(\pr_{a_3}\cdot\pr_{12})^{s_3-m_3-n+\g}\nonumber\\
&(\pr_{a_2}\cdot \pr_{a_3})^{\a}(\pr_{a_3}\cdot \pr_{a_1})^{\b}(\pr_{a_1}\cdot \pr_{a_2})^{\g}\label{hatT}\, ,\\
T(\a,\b,\g|m_1,m_2,m_3)=& \hspace{-15pt} T(\a,\b,\g|m_1,m_2,m_3)(\pr_{a_1}\cdot \pr_{1})^{m_1}(\pr_{a_2}\cdot \pr_{2})^{m_2}(\pr_{a_3}\cdot \pr_{3})^{m_3}\, .
\eea
We should add and subtract terms (\ref{defB}) and (\ref{defD}) to the (\ref{varL0}), and arrange the coefficients $B\ and\ D$ to satisfy equations, which ensure
\be
\begin{split}
&\d \cL+\D_2+\D_3-\D_2-\D_3=O((\pr_2\cdot\pr_{a_2})^n M(\vf^{s_2}(a_2)),(\pr_3\cdot\pr_{a_3})^k M(\vf^{s_3}(a_3))), \\
&n=0,1,...,s_2, \quad k=0,1,...,s_3.
\end{split}
\ee
The $\Delta$ terms that we add are to be combined with the terms that include d'Alembertian operator on the fields, to form equations of motion and their divergences. The tricky point here is that these terms themselves give rise to d'Alembertian operators, so they give back-reaction that is not possible to take into account order-by-order in general, but is possible to handle altogether simultaneously. This is the reason why we add and subtract corrections to all orders in divergences at the same time, and then get the following recursion relations on the coefficients:
\bea\label{cdb}
\nn
-(s_1-n+\a)C^{0,m_2,m_3}_{\a,\b,\g}+(s_3-m_3-n+\g+1)D^{0,m_2,m_3}_{\a,\b,\g}=B^{0,m_2+1,m_3}_{\a,\b,\g}(1-m_2),\\
\\
\nn
(s_1-n+\a)C^{0,m_2,m_3}_{\a,\b,\g}-(s_2-m_2-n+\b+1)B^{0,m_2,m_3}_{\a,\b,\g}=D^{0,m_2,m_3+1}_{\a,\b,\g}(1-m_3),\\
\label{cbd}
\eea
\bea\label{bcd}
(\g+1)[C^{0,m_2-1,m_3}_{\a,\b,\g+1}+C^{0,m_2,m_3}_{\a,\b,\g+1}]
+(\b+1)[C^{0,m_2,m_3-1}_{\a,\b+1,\g}-C^{0,m_2,m_3}_{\a,\b+1,\g}]\nonumber\\
+(\a+1)[B^{0,m_2,m_3-1}_{\a+1,\b,\g}+B^{0,m_2,m_3}_{\a+1,\b,\g}]
+(\a+1)[D^{0,m_2-1,m_3}_{\a+1,\b,\g}-D^{0,m_2,m_3}_{\a+1,\b,\g}]\nonumber\\
-(\g+1)B^{0,m_2,m_3}_{\a,\b,\g+1}+(\b+1)D^{0,m_2,m_3}_{\a,\b+1,\g}=0\,,
\eea
\be\label{bcd1}
(s_2-m_2-n+\b+1)B^{0,m_2,m_3}_{\a,\b,\g}-(s_3-m_3-n+\g+1)D^{0,m_2,m_3}_{\a,\b,\g}=-C^{1,m_2,m_3}_{\a,\b,\g}\,,
\ee
It is possible to check that the following choice for the coefficients $C$, $B$ and $D$
\be
\begin{split}
&C^{0,m_2,m_3}_{\a,\b,\g}=\frac{(-1)^{m_3}(1-m_2-m_3)}{\a!\b!\g!}\binom{s_2-n+\b}{m_2}\binom{s_3-n+\g}{m_3} \, ,\\
&B^{0,m_2,m_3}_{\a,\b,\g}=\frac{(-1)^{m_3+1}}{\a!\b!\g!}(s_1-n+\a)\binom{s_2-n+\b}{m_2-1}\binom{s_3-n+\g}{m_3}\, ,\\
&D^{0,m_2,m_3}_{\a,\b,\g}=\frac{(-1)^{m_3+1}}{\a!\b!\g!}(s_1-n+\a)\binom{s_2-n+\b}{m_2}\binom{s_3-n+\g}{m_3-1} \, ,\\
&C^{1,m_2,m_3}_{\a,\b,\g}=\frac{(-1)^{m_3}(m_2-m_3)}{\a!\b!\g!}(s_1-n+\a)\binom{s_2-n+\b}{m_2}\binom{s_3-n+\g}{m_3} \, ,\label{Sol}
\end{split}
\ee
solves the system \eqref{cdb}-\eqref{bcd1}.

This implies that it is always possible to compensate at least the deformation of one constraint adding the appropriate higher divergence tail. However, since the solution is not cyclic-symmetric with respect to the indices $m_i$, it cannot compensate the deformation of the constraints stemming from the variation of $\cL$ with respect to the second and third fields, {\it i.e.} the corresponding Lagrangian can be invariant with respect to gauge transformations of the fields $\vf^{(s_2)}$ and $\vf^{(s_3)}$, only with {\it non-vanishing deformations of transversality constraints} for these fields.  To summarise, we found that the introduction of couplings proportional to higher divergences via the solution \eqref{Sol} does not dispense with the need to deform \eqref{tdiff}.

Let us stress, however, that we are not proving the solution \eqref{Sol} to be unique, therefore the proof of the need for deformation is not yet complete. If we were able to find a set of coefficients $C$, $B$ and $D$ solving the system \eqref{cdb}-\eqref{bcd1} and cyclic-symmetric in the $m_i$ indices, then we could add the corresponding higher-divergence tail so as to render $\cL$ gauge invariant without any deformation of the constraints. We studied this possibility in one explicit case, the $3-2-1$ vertex: using Mathematica it is possible to see that the system \eqref{cdb}-\eqref{bcd1} does not admit any solution simultaneously compensating the deformations of the spin-2 and spin-3 constraints, thus allowing to conclude that corrections of the form \eqref{corr} are necessary in the most general case\footnote{In some special cases, the deformation can be avoided because the derivatives arrange in such a way to give rise to the equations of motion; for instance, in the spin-2 case the term $\pr\,\prd\prd h\sim \prd M$ may appear, or in the spin-3 case -- the term $\prd\prd\prd\vf\sim \prd M'$. However, this kind of mechanism strictly depends on the vertex under investigation and does not manifests itself in the general case.}.

\section{Spectra of partially reducible theories}\label{D}

In this section we analyse the spectra of three Maxwell-like theories on which we impose trace constraints of order higher than one. We wish to show that, as argued in Section \ref{sec: cubic_irred}, the lower-spin particles present in the fully reducible theory get truncated away from the spectrum. We shall resort to a light-cone analysis of the components, always assuming to choose a frame s.t. $p_+ \neq 0$.

\subsection*{Rank four: elimination of the scalar mode}\label{four}

Let us consider first the case of a rank--four tensor on which we impose a double-trace constraint. The corresponding theory is then characterised as follows:
\be
\begin{split}
& M_2 \, = \, M \, + \, \fr{4}{D \, (D + 2)} \, \h^{\, 2} \, \prd \prd \vf^{\, \pe} \, = \, 0 \, , \\
&\vf^{\, \pe \pe} \, = \, 0 \, , \\
& \prd \e \, = \, 0 \, .
\end{split}
\ee
From the divergence of the equations of motion we get
\be
\prd M_2 \, = \, - \, \pr \, \left(\prd \prd \vf \, - \, \fr{4}{D \, (D + 2)} \, \h \, \prd \prd \vf^{\, \pe} \right) \, = \, 0 \, ,
\ee
and thus, for normalisable modes
\be \label{ddf}
\prd \prd \vf \, - \, \fr{4}{D \, (D + 2)} \, \h \, \prd \prd \vf^{\, \pe} \, = \, 0 \, .
\ee
The trace of \eqref{ddf} allows to set to zero first $\prd \prd \vf^{\, \pe}$ and then, as a consequence, the whole of $\prd \prd \vf$. At this point we can set to zero the divergence of $\vf$ by a partial gauge fixing,
\be \label{C4}
\d \, \prd \vf \, = \, \Box \, \e \, \hskip 1cm \longrightarrow \hskip 1cm \prd \vf \, = \, 0 \, ,
\ee
since both $\prd \vf$ and $\e$ are transverse, rank--three tensors, thus obtaining the reduced Fierz system
\begin{align} \label{fierzred}
&\Box  \, \vf \, = \, 0\, ,& & & &\Box  \, \e\, = \, 0\, \, ,\nonumber\\
&\prd \vf \, = \,  0 \, ,& &\vf \, \sim \, \vf + \pr \, \e& &\prd \e\, = \,  0 \, , \\
&\vf^{\, \pe \pe} \,   = \, 0 \, , & & & \nonumber
\end{align}
that indeed describes two massless particles, one of spin $4$ and one of spin $2$.

\subsection*{Rank five: elimination of the vector mode}\label{five}

A slightly more complicated case obtains by considering a rank--five tensor subject to a double-trace condition. In this case, since we are aiming to truncate the vector of the Maxwell-like multiplet we should make sure that this is done in a gauge-invariant way and for this reason we need to also constrain the gauge parameter to be doubly traceless:
\be \label{C6}
\begin{split}
& M_2 \, = \, M \, + \, \fr{4}{(D + 4) \, (D + 2)} \, \h^{\, 2} \, \prd \prd \vf^{\, \pe} \, = \, 0 \, , \\
&\vf^{\, \pe \pe} \, = \, 0 \, , \\
& \prd \e \, = \, 0, \quad \e^{\, \pe \pe} \, = \, 0\,  .
\end{split}
\ee
Now the divergence of the equations of motion is not a pure gradient, and in order to investigate the corresponding cancellations we resort to a light-cone analysis in momentum space, assuming to choose a frame s.t. $p_{+} \neq 0$:
\be
p \cdot M_5 \, = \, - \, p \, p \cdot p \cdot \vf \, + \, \fr{4}{(D + 4) \, (D + 2)} \, \left(\h \, p \, p \cdot p \cdot \vf^{\, \pe}  \, 
+ \, \h^{\, 2} p \cdot p \cdot p \cdot \vf^{\, \pe} \right) \, = \, 0 \, .
\ee
The component analysis provides, to begin with, the following outcome:
\begin{align}
&+ \, + \, + \,  +  & &p \cdot p \cdot \vf_{\, + + +} \, = \, 0  \\
&+ \, + \, + \, i  &  &p \cdot p \cdot \vf_{\, + + i} \, = \, 0  \\
&+ \, + \, + \, -  & &p \cdot p \cdot \vf_{\, + + -} \, = \, - \, 2 \, \l \, p \cdot p \cdot \vf^{\,\pe}_{\, +}  \label{C11} \\
&+ \, + \ \ i \ \  j  &  &p \cdot p \cdot \vf_{\, + i j} \, = \, \l \, \h_{\, i j} \, p \cdot p \cdot \vf^{\,\pe}_{\, +}   \label{C12} \\
&+ \, + \ \ i \ \  -  &  &p \cdot p \cdot \vf_{\, + i -} \, = \, - \, \l \, \, \left(p \cdot p \cdot \vf^{\,\pe}_{\, i} \, - \, \label{C13}
\tfrac{p_i}{2 p_+} \, p \cdot p \cdot \vf^{\,\pe}_{\, +}\right)\, ,
\end{align}
where  $\l : = \tfrac{4}{(D + 4) (D + 2)}$. In particular, from the relation
\be
p \cdot p \cdot \vf^{\,\pe}_{\, +} \, = \, - \, 2 \, p \cdot p \cdot \vf_{\, + + -} \,  + \, p \cdot p \cdot \vf_{\, kk +} \, ,
\ee
putting together with \eqref{C11} and \eqref{C12} we obtain
\be \label{C15}
p \cdot p \cdot \vf^{\,\pe}_{\, +} \, = \, 0 \, \quad \longrightarrow \quad p \cdot p \cdot \vf_{\, + + -}\, = \, 0, \ \   p \cdot p \cdot \vf_{\, + i j} \, = \, 0 \, .
\ee
Continuing the analysis,
\begin{align}
&+  \ \ i \ \  j \ \ k  &  &p \cdot p \cdot \vf_{\, i j k } \, = \, \l \, \h_{\, (i j} \, p \cdot p \cdot \vf^{\,\pe}_{\, k)}  & & \label{C16} 
\end{align}
from which, since
\be \label{C17} 
p \cdot p \cdot \vf^{\,\pe}_{\, i} \, = \, - \, 2 \, p \cdot p \cdot \vf_{\, + - i } \,  + \, p \cdot p \cdot \vf_{\, kk i} \, ,
\ee
we obtain, putting together \eqref{C17}, \eqref{C16}, \eqref{C15} and \eqref{C13}
\be \label{C15}
p \cdot p \cdot \vf^{\,\pe}_{\, i} \, = \, 0 \, \quad \longrightarrow \quad p \cdot p \cdot \vf_{\, i j k}\, = \, 0, \ \   p \cdot p \cdot \vf_{\, + - i } \, = \, 0 \, .
\ee
For the remaining components we get
\begin{align}
&+ \, + \, - \, -  & &p \cdot p \cdot \vf_{\, + - -} \, + \, 2 \, \l \, p \cdot p \cdot \vf^{\, \pe}_{\, -} \, - \, \l \, p \cdot p \cdot p \cdot \vf^{\, \pe}= \, 0 \, , \\
&+ \ \  i \ \  j \ \ -  & &p \cdot p \cdot \vf_{\, i j -} \, + \, \l \, \h_{\, ij} \, p \cdot p \cdot \vf^{\, \pe}_{\, -} \, - \, 
\tfrac{\l}{p_+} \, p \cdot p \cdot p \cdot \vf^{\, \pe}= \, 0 \, ,  \\
&+ \, i \, - \ \ -  & &p \cdot p \cdot \vf_{\, i - -} \, +  \,  \tfrac{p_i}{p_+}\, p \cdot p \cdot \vf_{\, + - -} \, + \, 
2 \,  \l \, \tfrac{p_i}{p_+} \, p \cdot p \cdot \vf^{\, \pe}_{\, -} \, = \, 0 \, ,\\
&+ \, - \,  - \,  -  & &p \cdot p \cdot \vf_{\, - - -} \, +  \,  3 \, \tfrac{p_-}{p_+}\, p \cdot p \cdot \vf_{\, + - -} \, + \, 
6 \,  \l \, \tfrac{p_-}{p_+} \, p \cdot p \cdot \vf^{\, \pe}_{\, -} \, = \, 0 \,  .
\end{align}
In the first two conditions we can substitute
\be
\begin{split}
& p \cdot p \cdot p \cdot \vf^{\, \pe} \, = \, 2 \, p_+ \, p \cdot p \cdot \vf_{\, + - -} \,  - \, p_+ \, p \cdot p \cdot \vf_{\, k k -} \, ,\\
& p \cdot p \cdot \vf^{\, \pe}_{\, -} \, = \, - \, 2 \, p \cdot p \cdot \vf_{\, + - -} \,  + \, p \cdot p \cdot \vf_{\, k k -} \, .
\end{split}
\ee
Tracing the second equation and solving the system with the first one obtains
\be
p \cdot p \cdot \vf_{\, + - -} \, = \, 0 \, ,\quad \quad p \cdot p \cdot \vf_{\, k k  -} \, = \, 0 \, ,
\ee
from which all leftover components of the double divergence of $\vf$ can be seen to vanish. Altogether one finds
\be
p \cdot p \cdot \vf_{\, \m \n \r} \, = \, 0 \, .
\ee
The gauge dependent part of the divergence, in its turn, can be set to zero by a partial gauge fixing as in \eqref{C4}, observing that in the present case both $\prd \vf$ and $\e$ are transverse and doubly traceless. In this way we prove equivalence of \eqref{C6} and \eqref{fierzred}, where now all tensors refer to the rank--five case and $\e^{\, \pe \pe} = 0$ is also understood, thus showing that \eqref{C6} propagates only the particles of spin $5$ and spin $3$, with the vector eliminated from the spectrum.

\subsection*{Rank six: elimination of the spin-$0$ and of the spin-$2$ modes}\label{six}

As our last example we would like to consider a case where trace constraints are supposed to cut half of the spectrum of the original theory. To this end we impose on a rank--six tensor the following set of conditions:
\be \label{C6}
\begin{split}
& M_2 \, = \, M \, + \, \fr{4}{(D+6) \, (D + 4)} \, \h^{\, 2} \, \prd \prd \vf^{\, \pe} \, = \, 0 \, , \\
&\vf^{\, \pe \pe} \, = \, 0 \, , \\
& \prd \e \, = \, 0, \quad \e^{\, \pe \pe} \, = \, 0\,  .
\end{split}
\ee
Once again, the main issue at stake is to show that the equations of motion are strong enough to impose the vanishing of the double divergence. As discussed above, the transverse part of the divergence can be eliminated by a partial gauge fixing. Thus, we consider the strongest condition emerging on $\prd \prd \vf$, enforced by the divergence of the equations of motion:
\be \label{M6}
p \cdot M_2 \, = \, - \, p \, p \cdot p \cdot \vf \, + \, \fr{4}{(D + 6) \, (D + 4)} \, \left(\h \, p \, p \cdot p \cdot \vf^{\, \pe}  \, 
+ \, \h^{\, 2} p \cdot p \cdot p \cdot \vf^{\, \pe} \right) \, = \, 0 \, .
\ee
Let us go through the component analysis  of \eqref{M6} ($\m : = \tfrac{(D + 6) \, (D + 4)}{4}$):
\begin{align}
&+ \, + \, + \,  +  \, + \, & & p \cdot p \cdot \vf_{\, + + + +} \, = \, 0  \\
&+ \, + \, + \, +\ \,  i  &  & p \cdot p \cdot \vf_{\, + + + \, i} \, = \, 0  \\ 
&+ \, + \, + \, + \, -   &  & (\m \, - \, 6) \, p \cdot p \cdot \vf_{\, + + + -} \, = \, - \, 3 \,  \, p \cdot p \cdot \vf_{\, + + k k}   \label{D29} \\
&+ \, + \, + \ \, i \ \, j   &  & \m \,  p \cdot p \cdot \vf_{\, + + \, i j} \, = \, \d_{\, i j} \, (- \, 2 \, p \cdot p \cdot \vf_{\, + + + -} \, 
+ \, p \cdot p \cdot \vf_{\, + + k k})   \label{D30} \, ,\\
&+ \, + \, + \, - \ \, i   &  & (\m \, - \, 4) \, p \cdot p \cdot \vf_{\, + + - \, i} \, = \, - \, 2 \,  \, p \cdot p \cdot \vf_{\, + i k k}   \label{D31} \\
&+ \, + \ \, i \ \, j \ \, k  &  &  \m \,  p \cdot p \cdot \vf_{\, + \, i \, j\, k} \, = \, \d_{\, (i j} \, (- \, 2 \, p \cdot p \cdot \vf_{\,k) + + -} \, 
+ \, p \cdot p \cdot \vf_{\, k) \, \ell \ell - })   \label{D32} \, ,
\end{align}
where in particular, tracing over the transverse indices in \eqref{D30} and solving together with \eqref{D29} one obtains $p \cdot p \cdot \vf_{\, + + + -}  = 0$ and $p \cdot p \cdot \vf_{\, + + k k} = 0$ and, inserting back in \eqref{D30}, $p \cdot p \cdot \vf_{\, + + \, i j} = 0$. Similarly, combining the trace of \eqref{D32} with \eqref{D31} one obtains eventually $p \cdot p \cdot \vf_{\, + + - \, i}$ and $p \cdot p \cdot \vf_{\, + \, i \, j\, k}$.  The following equations form another coupled system 
\begin{align}
&+ \, + \, + \, - \, -  &  &  (\m \, - \, 16) \,   p \cdot p \cdot \vf_{\, + \, + \, -\, -} \, = \, - \, 8 \, p \cdot p \cdot \vf_{\,+ - k k } \, ,\label{D33} \\
&+ \, + \, - \ \, i \ \, j   &  &  (\m \, - \, 2)\, p \cdot p \cdot \vf_{\, + \, - \, i\, j} \, = \, 2 \, \d_{\, i j} \, p \cdot p \cdot \vf_{\, + - k k } -  
\,  4 \, \d_{\, i j} \, p \cdot p \cdot \vf_{\, + + - -} \nonumber \\ 
& & &  \, -  \, p \cdot p \cdot \vf_{\, i j k k}   \label{D34}  \, , \\
&+ \ \, i \ \, j \ \, k \ \, \ell  &  &  \m \, p \cdot p \cdot \vf_{\, i\, j \, k \, \ell} \, = \, \d_{\, (i j} \, p \cdot p \cdot \vf^{\, \pe}_{\, k \ell)} \, ,  \label{D35}
\end{align}
whose solution requires to also take the double trace condition into account:
\be \label{D36}
\vf^{\, \pe \pe}{}_{\, \m \n} \, = \, 4 \, \vf_{\, + + - - \, \m \n }  \, - \, 4 \, \vf_{\, + - \ell \ell \, \m \n }  \, + \,  \vf_{\,k k  \ell \ell \, \m \n } \, = \, 0 \, .
\ee
To begin with, one can combine the trace of \eqref{D34} with \eqref{D33} and \eqref{D36} to get $p \cdot p \cdot \vf_{\, + + - -} = 0$ and $p \cdot p \cdot \vf_{\,+ - k k } = 0$. However, in \eqref{D34} one is still left with the full traceless part (in the transverse indices) of 
$p \cdot p \cdot \vf_{\, + \, - \, i\, j}$.  To check that it vanishes, one has to combine \eqref{D34} with the trace of \eqref{D35} and again \eqref{D36}, to eventually recover  $p \cdot p \cdot \vf_{\, + \, - \, i\, j} = 0$ together with $p \cdot p \cdot \vf_{\, i\, j \, k \, \ell} = 0$.
The reasoning is similar for the remaining components. In particular from the relations
\begin{align}
&+ \, +  \, - \, - \ \, i  &  &  (\m \, - \, 8) \,   p \cdot p \cdot \vf_{\, + \, i \, -\, -} \, = \, - \, 4 \, p \cdot p \cdot \vf_{\, k k i -}
\, + \, 2 \, \fr{p_k}{p_+} \,  p \cdot p \cdot \vf^{\, \pe}_{\, k i}   \label{D37} \\
&+ \, +  \, - \, - \, - &  &  (\m \, - \, 36) \,   p \cdot p \cdot \vf_{\, + \, - \, -\, -} \, = \, - \, 9 \, p \cdot p \cdot \vf_{\, k k - -} \nonumber \\
& & & \, - \, 6 \, \fr{p_k}{p_+} \, \left(2\, p \cdot p \cdot \vf_{\, + - - k} \, + \,  p \cdot p \cdot \vf_{\, k \ell \ell - }\right)  \label{D38} \\
&+ \, - \ \, i \ \, j \ \, k &  & \m \, p \cdot p \cdot \vf_{\, i \, j \, k \, -} \, = \, 2 \,  \d_{\, (i j} \, p \cdot p \cdot \vf^{\, \pe}_{\, k) -}   \label{D39} \\
&+ \, - \, - \ \, i \ \, j  &  & \m \, p \cdot p \cdot \vf_{\, i \, j \, - \, -} \, = \, 3 \,  \d_{\, i j} \, p \cdot p \cdot \vf^{\, \pe}_{\, k k - -} 
\, - \, 6 \,  \d_{\, i j} \, p \cdot p \cdot \vf^{\, \pe}_{\, + - - -} \, ,\label{D40}
\end{align}
one has to first remove the traces in the transverse indices, combining \eqref{D39} with \eqref{D37} and \eqref{D40} with \eqref{D38} to then get
$ p \cdot p \cdot \vf_{\, + \, i \, -\, -} = 0$, $p \cdot p \cdot \vf_{\, i \, j \, k \, -} = 0$, $ p \cdot p \cdot \vf^{\, \pe}_{\, + - - -} = 0$ and $p \cdot p \cdot \vf_{\, i \, j \, - \, -} = 0$.  At this stage actually one has shown that 
\be
p \cdot p \cdot \vf^{\, \pe}{}_{\m \n } \, = \, 0 \, ,
\ee
which makes immediate to eliminate the last components, 
\begin{align}
&+ \, -  \, - \, - \ \, i  &  &  p \cdot p \cdot \vf_{\, i - - - } \, = \, 0 \, ,  \label{D42} \\
&+ \, -  \, - \, - \, - &  & p \cdot p \cdot \vf_{\, - - - -} \, = \, 0 \, ,\label{D43} 
\end{align}
thus completing the proof that $\prd \prd \vf = 0$ on shell.

\bibliographystyle{JHEP}

\end{document}